\documentclass[lettersize,journal]{IEEEtran}
\usepackage{amsmath,amsfonts}
\usepackage{algorithmic}
\usepackage{algorithm}
\usepackage{array}
\usepackage[caption=false,font=footnotesize]{subfig}
\usepackage{textcomp}
\usepackage{stfloats}
\usepackage{url}
\usepackage{verbatim}
\usepackage{graphicx}
\usepackage{cite}
\usepackage{enumerate}
\usepackage{epstopdf}
\usepackage{booktabs}
\usepackage{float}
\usepackage{color}
\usepackage{multirow}
\usepackage{tabularx}
\usepackage{makecell}
\begin{document}
\title{High-Resolution Multi-Target DOA Estimation for Resonant Beam Systems}

\author{Guangkun Zhang, Mingqing Liu, Wen Fang, Mingliang Xiong,\\Yunfeng Bai and Qingwen Liu
\thanks{Guangkun Zhang and Qingwen Liu are with Shanghai Research Institute for Autonomous Intelligent Systems, Tongji University, Shanghai 201804, China. (e-mail:2210993@tongji.edu.cn; qliu@tongji.edu.cn;)}
\thanks{Yunfeng Bai, Mingliang Xiong and Qingwen Liu are with the School of Computer Science and Technology,
Tongji University, Shanghai 201804, China. (e-mail: baiyf@tongji.edu.cn; mlx@tongji.edu.cn; qliu@tongji.edu.cn)}
\thanks{Wen Fang and Mingqing Liu are with the College of Electronics and Information Engineering, Tongji University, Shanghai 201804, China. (e-mail: wen.fang@tongji.edu.cn; clare@tongji.edu.cn)}
}

\maketitle
\begin{abstract}
Direction of arrival (DOA) estimation technology offers a promising solution to address the sensing and positioning demands of Internet of Things (IoT) devices. Optical resonant beam systems (RBS), owing to their inherent characteristics of self-alignment, self-established energy focusing, and passive target sensing, make them naturally suited for {\color{blue}DOA} estimation in IoT scenarios. However, RBS suffer from limited angular resolution and a narrow field of view (FoV) in multi-target environments. To overcome these limitations, this paper proposes a high-resolution wide-field-of-view resonant beam DOA estimation system (RB-HWDOA). The RB-HWDOA integrates an optical spectrum-based DOA estimation algorithm (OSB-DOA), which leverages amplitude information in the two-dimensional Fourier spectrum of the resonant beam, {\color{blue}overcoming the resolution limit imposed by the beam size in spatial-domain methods}. Furthermore, we designed a {\color{blue}telescope} modulation (TM) structure to correct phase and direction mismatches, enabling a multi-Tx framework that focuses beams onto a common sensing module, thereby extending the effective FoV. Combined with the OSB-DOA algorithm, this design supports high-resolution DOA estimation for {\color{blue}multiple targets simultaneously over a wide FoV}. Simulation results show that OSB-DOA resolves angular separations down to $0.1^{\circ}$ across multiple resonant beams, remains robust under noise, {\color{blue}and the TM architecture enables multi-Tx integration for wide-FoV coverage}, making RB-HWDOA a scalable and efficient solution for passive multi-target DOA estimation in complex IoT environments.
\end{abstract}

\begin{IEEEkeywords}
Resonant beam, direction of arrival, {\color{blue}multiple targets}, wide field of view, high-resolution.
\end{IEEEkeywords}

\section{Introduction}

{\color{blue}DOA} estimation is a fundamental technique in array signal processing and plays a significant role in wireless communication and sensing. For instance, in smart antenna systems, DOA estimation facilitates adaptive beamforming, which enhances communication capacity while effectively reducing power consumption and improving overall efficiency \cite{ref2}. As illustrated in Fig.~1, DOA estimation plays a pivotal role in facilitating precise positioning and communication for {\color{blue}multiple targets}, thereby enhancing the performance of various IoT-driven applications, including autonomous driving, smart cities, and mobile communication systems. Given its critical function and growing demand in IoT applications, recent research has begun to explore new system architectures and efficient methods aimed at enhancing DOA estimation performance while reducing implementation complexity and power consumption \cite{ref4, ref5}.

\begin{figure}[!t]
\centering
\includegraphics[width=3in]{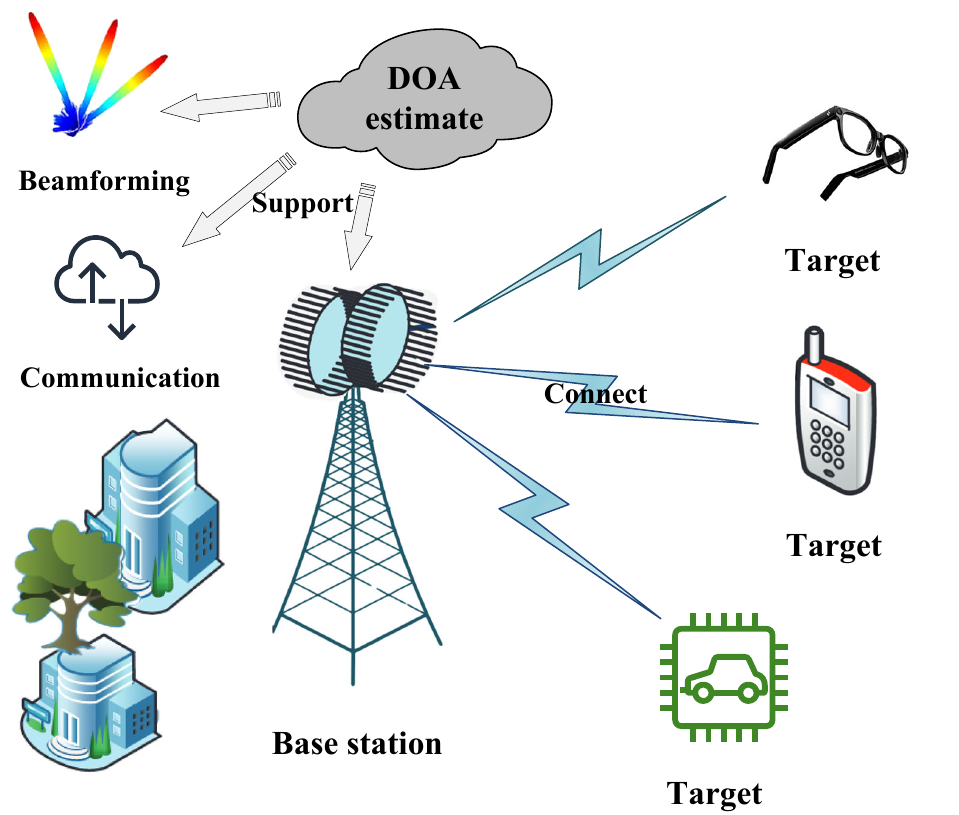}
\caption{DOA-based service application scenarios.}
\label{fig_1}
\end{figure}

Conventional radio-frequency (RF) {\color{blue}DOA} estimation algorithms are predominantly based on multi-channel array signal processing. For instance, the minimum variance distortionless response (MVDR) {\color{blue}algorithm} \cite{ref5} estimates directions by minimizing output power while preserving unity gain in the look direction. To achieve higher angular resolution, subspace-based methods such as multiple signal classification (MUSIC) \cite{ref6} and estimation of signal parameters via rotational invariance techniques (ESPRIT) \cite{ref7} employ eigenvalue decomposition of the signal covariance matrix to separate signal and noise subspaces, thereby enabling high-resolution DOA estimation. {\color{blue}However, these multiantenna systems generally require multiple RF front-end chains, each comprising amplification, downconversion, filtering, analog-to-digital conversion, and digital baseband processing \cite{ref8,ref9}.} This architecture results in substantial hardware complexity and power consumption, rendering it less suitable for large-scale, low-cost, and energy-efficient IoT applications. Recently, programmable metasurfaces have emerged as a promising alternative for DOA estimation \cite{ref10}. Through precise control over the electromagnetic response of subwavelength unit cells, metasurfaces enable dynamic manipulation of phase, amplitude, and polarization. They have been extensively utilized in applications such as radar absorbers, orbital angular momentum beams, and computational imaging \cite{ref11,ref12}. Further studies have proposed multi-target DOA estimation techniques using spatiotemporally modulated metasurfaces. By exploiting time-delay effects, these methods can accomplish directional sensing with only a single receive sensor, presenting opportunities for compact and low-cost implementations \cite{ref13}.

Beyond microwave and millimeter-wave systems, {\color{blue}DOA} estimation in the optical domain has garnered growing interest, owing to the capability of optical wireless (OW) systems to deliver high spatial resolution and robustness against RF interference \cite{ref14}. Typical optical DOA methods determine the incident direction of light signals by leveraging imaging geometry or optical phase analysis \cite{ref15,ref16}. For example, some approaches utilize geometric imaging based on optical beacons to measure the angular relationship between the beacon and receiver, while laser systems employ phase-sensing mechanisms to realize compact and high-precision angle measurements \cite{ref17}. However, these techniques often necessitate precise optical alignment or are applicable primarily to active targets. In applications involving passive {\color{blue}targets} (especially those dependent on weak reflected or scattered optical signals){\color{blue},} conventional optical DOA methods exhibit limited effectiveness. This underscores the need for new, robust approaches capable of operating under challenging conditions such as low signal-to-noise ratio (SNR) and multipath propagation.

RBS have recently emerged as a promising approach for DOA estimation, demonstrating notable performance in both optical and RF regimes \cite{ref18,ref19,ref20}. Originally developed for optical wireless power transfer in \cite{ref21}, RBS form a resonant cavity between transmitter (Tx) and receiver (Rx) using cat-eye retroreflectors. Within this cavity, laser beams are generated and sustained through repeated reflection and amplification via a gain medium. This process suppresses scattered radiation and {\color{blue}produces} highly collimated, energy-focused beams \cite{ref19}. An analogous implementation in the RF domain employs mutually reflective antenna arrays to emulate the phase-confinement behavior of an optical cavity, thereby achieving adaptive beamforming without explicit channel estimation \cite{ref22}. Owing to its inherent properties, including passive operation at the target, intrinsic beam collimation, self-alignment, self-tracking capability, and energy focusing, RBS offers a compelling architecture for DOA estimation in IoT applications \cite{ref21}.

\begin{figure}[!t]
\centering
\includegraphics[width=3in]{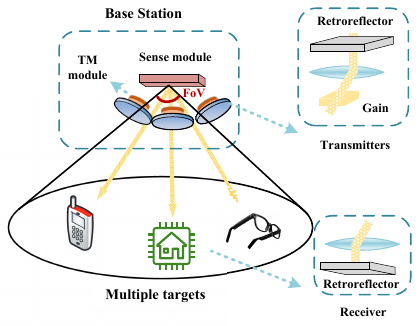}
\caption{{\color{blue}Integrated multi-Tx resonant beam base station (BS) for FoV expansion.}}
\label{fig_2}
\end{figure}

Optical RBS offers a structurally simpler {\color{blue}architecture, making} it more amenable to integration into lightweight IoT devices. In optical RBS, two primary methodologies are employed for DOA estimation. The first approach{\color{blue},} referred to as the optical spot centroid-based DOA estimation algorithm (OSCB-DOA){\color{blue},} utilizes the geometric configuration of the cat-eye retroreflector to estimate DOA \cite{ref18,ref23,ref24}. This method offers computational efficiency due to its linear complexity, though its angular resolution is limited in multi-target scenarios. The second approach adapts classical RF DOA estimation techniques to optical signals, termed the {\color{blue}optical wavefront phase-based} DOA estimation algorithm (OWPB-DOA) \cite{ref17}. While OWPB-DOA achieves higher resolution, it requires high-precision sensor sampling and entails substantially higher computational complexity. Furthermore, the RBS architecture exhibits a limited FoV during DOA estimation. Despite the compact nature of optical components, individual Tx and Rx are typically constrained to an FoV of approximately $8^{\circ}$. This restriction impedes the deployment of optical RBS in wide-area or large-scale application scenarios.

To address these limitations, this paper proposes a high-resolution wide-field-of-view resonant beam DOA estimation framework, termed RB-HWDOA. We introduce the optical spectrum-based DOA estimation algorithm (OSB-DOA) to estimate DOA via spatial spectrum analysis. {\color{blue}This algorithm leverages} the phase-sensitive oscillation of resonant beams to overcome the beam width-limited resolution of amplitude-only spatial distribution, achieving high-resolution in multi-target scenarios. Furthermore, we developed a multi-Tx framework to extend the effective FoV and design a telescope modulation (TM) structure for Tx to implement this architecture. This TM structure actively corrects phase and directional misalignments, allowing multiple Tx beams to focus onto a common sensing module. As illustrated in Fig. 2, this scheme ultimately achieves wide-angle coverage. The contributions of this paper can be {\color{blue}summarized} as follows.

\begin{enumerate}
\item We propose an OSB-DOA estimation algorithm that leverages the phase-sensitive oscillation of resonant beams to overcome the beam width-limited resolution of amplitude-only spatial distributions. By exploiting amplitude information encoded in the two-dimensional Fourier spectrum of the complex field, the algorithm enables angular discrimination with a minimum resolvable angle of $0.1^{\circ}$, while maintaining low computational complexity and strong noise robustness for accurate {\color{blue}identification of multiple targets}.
\item We propose a multi-Tx framework for RBS that expands the effective FoV by distributing Tx over a spherical surface and synchronizing their resonant beams. To realize this, we design a TM structure that focuses beams arriving from different directions onto a common sensing module, actively {\color{blue}correcting} phase and directional mismatches. Moreover, we verify through simulations that the proposed design achieves high-resolution DOA estimation for {\color{blue}multiple targets}, lightweight implementation, and scalable FoV enhancement.
\end{enumerate}

The remainder of this paper is organized as follows. Section II presents the system architecture and operating principle of the RB-HWDOA. Section III introduces the mathematical model and simulation tools used for the RB-HWDOA system. Section IV provides the simulation results derived from the analytical model. Finally, Section V summarizes the contributions of this work and outlines directions for future research.

\section{System Overview}

\begin{figure}[!t]
\centering
\includegraphics[width=3.4in]{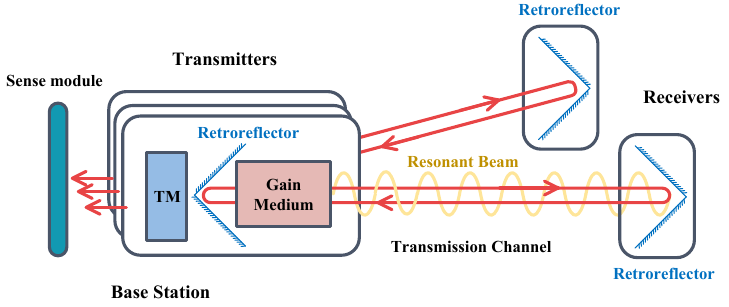}
\caption{The architecture of the RB-HWDOA system.}
\label{fig_3}
\end{figure}

\begin{figure}[!t]
\centering
\includegraphics[width=3.3in]{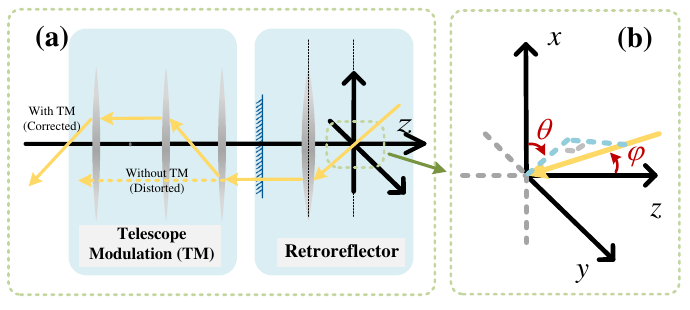}
\caption{{\color{blue}Functional schematic of the TM structure: (a) optical path illustrating how the TM module corrects the propagation direction of the resonant beam after the retroreflector; (b) definition of the elevation angle $\phi$ and azimuth angle $\theta$. The detailed lens configuration and parameters are given in Fig.~7.}}
\label{fig_4}
\end{figure}

The architecture of the RB-HWDOA system, illustrated in Fig.~3, comprises two primary components: the base station and the Rx. The base station includes Tx and a sensing module, which incorporates retroreflectors, gain media, and a TM module. {\color{blue}The} base station is responsible for performing DOA sensing and supplying the energy necessary to generate resonant {\color{blue}beams}. The Rx, which consists of a single retroreflector, is mounted on the target. Together, the Tx and Rx form the infrastructure for establishing resonant {\color{blue}beams}, with each Tx-Rx pair capable of independently generating a resonant beam to form a DOA information transmission link. This capability enables the RB-HWDOA system to perform DOA estimation for {\color{blue}multiple targets} simultaneously. 

{\color{blue}The convex lens inside the cat-eye retroreflector distorts the propagation direction of the resonant beam during single-pass transmission. To compensate for this directional distortion, we introduce a TM module composed of three additional convex lenses. As shown in Fig.~4(a), without the TM module the beam exits the retroreflector along a distorted direction, whereas with the TM module the beam's original propagation direction is restored and refocused toward the sensing plane. Consequently, a resonant beam incident from an arbitrary direction, defined by elevation angle $\phi$ and azimuth angle $\theta$ in Fig.~4(b), traverses the retroreflector and the TM module while preserving its original propagation direction.}

{\color{blue}As shown in Fig.~5, after passing through the retroreflector lens (focal length $f_1$) and the TM module, the corrected beam refocuses at the back focal point of the last TM lens (focal length $f_2$) with zero lateral displacement (derived in Section~III-D). Based on this property, the Tx units are distributed on a sphere of radius $f_2$ centered on the sensing module. Since each Tx's output lens has the same focal length $f_2$, all back focal points coincide at the sphere center, so the corrected beams from all Tx units naturally converge at a single common point. This point serves as the origin of the global DOA coordinate system and the location of the sensing module. This design achieves multi-Tx beam convergence purely through optical geometry, without active alignment or algorithmic calibration, enabling scalable FoV expansion with a compact form factor.}

{\color{blue}The resonant beams from all Tx units are focused onto the central wavefront sensor within the sensing module, which simultaneously captures both the phase and amplitude of the incident wavefront. Since the TM module preserves the beam's original propagation direction, the DOA measured at the sensing plane faithfully represents the true incident angle of the resonant beam, regardless of the Tx unit's own tilt. This property allows each Tx to be arbitrarily oriented while the system seamlessly integrates multiple Tx units for FoV expansion. For a target in overlapping Tx coverage, multiple links may be established. After TM correction, these links are mapped onto the common sensing plane with the same physical DOA; any residual peak offset is only the small parallax caused by the finite Tx baseline. In practice, this offset is typically below the OSB-DOA resolution, so the peaks appear merged as a single dominant peak, enabling consistent and robust DOA estimation without additional inter-Tx/boundary fusion.}

In multi-target scenarios, resonant beams with different incident angles may originate from the same or different Tx units. {\color{blue}The OSB-DOA algorithm estimates the DOA based on the principle that the displacement of the spectral peak from the center of the two-dimensional frequency domain is linearly determined by the incident angles. Combined with the linearity of the Fourier transform and the collimation characteristics of the resonant beam, this principle enables high-resolution DOA estimation for multiple resonant beams simultaneously.}

{\color{blue}Notably, the resonant beam link does not require strict alignment between the Tx and Rx. Owing to the self-alignment property of the resonant cavity, a stable resonant beam can be established as long as the Rx retroreflector falls within the FoV of the Tx (approximately $\pm 7^{\circ}$ per individual Tx unit, as characterized in Section~IV-C). This angular tolerance significantly relaxes the deployment constraints for practical IoT scenarios.}

\begin{figure}[!t]
\centering
\includegraphics[width=3.2in]{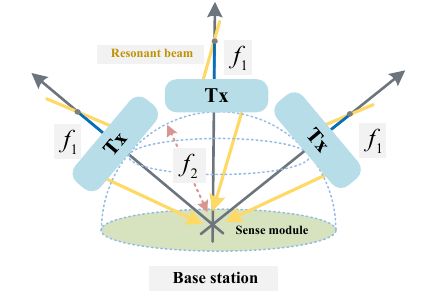}
\caption{{\color{blue}Schematic diagram of Tx distribution and resonant beam correction, where $f_1$ is the focal length of the retroreflector lens and $f_2$ is the focal length of the last lens of the TM module. The Tx units are distributed on a sphere of radius $f_2$ centered on the sensing module, so that all corrected beams naturally converge at the sphere center for DOA estimation.}}
\label{fig_5}
\end{figure}

\section{Mathematical Model}
{\color{blue}This section presents the mathematical model of the RB-HWDOA system. We first model the resonant beam formation via the Fox-Li iterative algorithm (Section~III-A), then derive the spectral-shift-based DOA estimation principle of the OSB-DOA method (Section~III-B) and its spatial frequency resolution limits (Section~III-C). Section~III-D analyzes the directional distortion of the retroreflector and the TM correction mechanism. Sections~III-E and III-F address the SNR model and the high-resolution mechanism, respectively.}

\subsection{Intracavity Field Formation and Stabilization}

In an optical resonator, the formation of resonant beams originates from the repeated diffraction and feedback cycles of the optical field under finite-aperture constraints. The Fox-Li algorithm \cite{ref18} simulates the generation of resonant beams through numerical iterations {\color{blue}as shown in Fig.~6}. 

{\color{blue}To analyze the beam propagation characteristics, we establish a Cartesian coordinate system $(x, y, z)$, where the $z$-axis aligns with the optical axis of the resonator, representing the longitudinal propagation direction. The coordinates $x$ and $y$ denote the transverse positions perpendicular to the optical axis, as shown in Fig.~6. Consequently, the optical field distribution on any given transverse plane is represented by its complex amplitude $E(x, y)$.} The beam formation can be expressed as the iterative application of a sequence of operators to the initial field distribution $E_0(x, y)$, following the rule:
\begin{equation}
E_{n+1}(x, y) = \Gamma_{\text{round}}[E_n(x, y)],
\end{equation}

where $\Gamma_{\text{round}}$ denotes the single round-trip operator. {\color{blue}Taking the transmitter mirror as the reference plane, $\Gamma_{\text{round}}$ represents the cascaded action of all optical elements in the RBS on the field over one complete circulation. As shown by the RBS architecture in Fig.~3 and the retroreflector/TM structure in Fig.~4, these optical elements determine how the operators in $\Gamma_{\text{round}}$ are concatenated, and the round-trip operator can be written as the ordered product of the propagation and modulation operators:
\begin{equation}
\begin{aligned}
\Gamma_{\text{round}} &=  \Gamma_{\text{M}}\Gamma_{\text{FS}}\Gamma_{\text{L}}\Gamma_{\text{FS}}\Gamma_{\text{G}}\Gamma_{\text{FS}}\Gamma_{\text{L}}\Gamma_{\text{FS}}\Gamma_{\text{M}} \\
& \cdot \Gamma_{\text{FS}}\Gamma_{\text{L}}\Gamma_{\text{FS}}\Gamma_{\text{G}}\Gamma_{\text{FS}}\Gamma_{\text{L}}\Gamma_{\text{FS}}\Gamma_{\text{M}}
\end{aligned} ,
\end{equation}
where the operators are applied from right to left. The second line corresponds to the forward propagation path from the Tx to the Rx, while the first line corresponds to the return path. Specifically, the sequence includes mirror reflection $\Gamma_{\text{M}}$, lens phase shift $\Gamma_{\text{L}}$, gain medium $\Gamma_{\text{G}}$, and free-space propagation $\Gamma_{\text{FS}}$.} 

\begin{figure}[!t]
\centering
\includegraphics[width=3in]{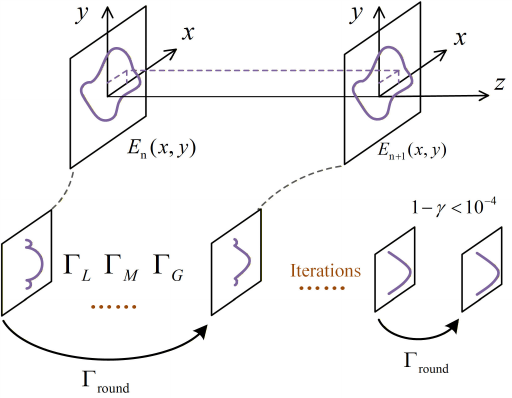}
\caption{{\color{blue}The process of generating self-reproducing modes of the resonant beam.}}
\label{fig_6}
\end{figure}

The mirror reflection operator $\Gamma_{\text{M}}$ models the aperture-limited reflection of the beam and is defined as:
\begin{equation}
\Gamma_{\text{M}}[E(x, y)] = E(x, y) \cdot M(x, y),
\end{equation}
where $M(x, y)$ is the mirror aperture function. Inside the aperture region i.e. $x^2 + y^2 \leq {\color{blue}r_m}^2$, $M(x, y) = 1$; otherwise, $M(x, y) = 0$. Here, ${\color{blue}r_m}$ is the mirror radius.

The lens operator $\Gamma_{\text{L}}$ introduces quadratic phase modulation to simulate the focusing effect of the lens, expressed as:
\begin{equation}
\Gamma_{\text{L}}[E(x, y)] = E(x, y) \cdot L(x, y),
\end{equation}
with the phase function:
\begin{equation}
L(x, y) =
\begin{cases}
\exp\left[-i \frac{\pi}{\lambda f}(x^2 + y^2)\right], & x^2 + y^2 \leq r_l^2 \\
0, & \text{otherwise}
\end{cases}
\end{equation}
where $r_l$ is the lens radius, $\lambda$ is the wavelength, and $f$ is the focal length.

The gain medium operator $\Gamma_{\text{G}}$ models the spatial modulation of the active medium and is defined as:
\begin{equation}
\Gamma_{\text{G}}[E(x, y)] = E(x, y) \cdot G(x, y),
\end{equation}
where, within the gain medium region $(x^2 + y^2 \leq r_g^2)$, $G(x, y) = \exp\left[\tfrac{g(x,y){\color{blue}L_g}}{2}\right]$; otherwise, $G(x, y) = 0$. Here, $g$ is the gain coefficient, ${\color{blue}L_g}$ is the medium length, and $r_g$ is the radius of the gain medium.

The free-space propagation operator $\Gamma_{\text{FS}}$ uses the angular spectrum method to simulate diffraction during free-space propagation:
\begin{equation}
\begin{aligned}
\Gamma_{\mathrm{FS}}[E(x,y);d]
&= \mathcal{F}^{-1}\Big\{ \mathcal{F}[E(x,y)]\\
&\qquad \cdot S(k_x,k_y,s_x,s_y)\\
&\qquad \cdot H(k_x,k_y,d) \Big\},
\end{aligned}
\end{equation}
where $\mathcal{F}$ and $\mathcal{F}^{-1}$ denote the two-dimensional Fourier transform and its inverse, $S(k_x, k_y, s_x, s_y)$ represents the propagation shift of the resonant beam, and $H(k_x, k_y, d)$ is the free-space transfer function. Here, $k_x$ and $k_y$ are {\color{blue}angular} spatial frequencies, the transfer function is defined in the spatial frequency domain {\color{blue} where $s_x$ and $s_y$ denote the transverse spatial displacements, and $d$ denotes the propagation distance}.

The iteration starts from an arbitrary initial field $E_0(x, y)$ and gradually converges to a stable solution after repeated updates. Convergence is achieved when the criterion is satisfied \cite{ref19}:
{\color{blue}\begin{equation}
\gamma = \frac{\|E_{n+1}(x, y)\|}{  \|E_n(x, y)\|}, \quad \text{with} \quad
1-\gamma< 10^{-4},
\end{equation}}
where $\|\cdot\|$ denotes the Euclidean norm (i.e., for a complex field $E(x,y)$, $\|E(x,y)\| = \sqrt{\iint |E(x,y)|^2 \mathrm{d}x\mathrm{d}y}$), and $\gamma$ denotes the round-trip eigenvalue, which approaches unity, indicating the stabilization of the dominant resonant mode with low loss. The multiple iterations of the {\color{blue}Fox-Li} algorithm describe the formation of resonant beams as a diffraction-selection mechanism. The finite-aperture mirror reflection operator $\Gamma_{\text{M}}$ imposes a hard boundary condition on the beam, which in the spatial domain is equivalent to multiplying the field by a circular window function

\begin{equation}
w(x, y) = \text{circ}\!\left(\frac{\sqrt{x^2 + y^2}}{{\color{blue}r_m}}\right),
\end{equation}

where $\text{circ}(r)$ is the unit circular window function, defined as $\text{circ}(r) = 1$ for $r \leq 1$ and $\text{circ}(r) = 0$ otherwise, and ${\color{blue}r_m}$ is the {\color{blue}mirror aperture} radius. \textcolor{blue}{According to the convolution theorem, this spatial multiplication corresponds to the convolution of the field's spectrum with the Fourier transform of the aperture function. This operation introduces diffraction losses, effectively suppressing high-order modes with larger spatial frequencies during iterative propagation, thereby acting as a spatial low-pass filter.}

{\color{blue}
\subsection{DOA Estimation via Spatial Spectral Shift}
This subsection establishes the mathematical model of the OSB-DOA. We demonstrate that a tilt of the resonator relative to the optical axis induces a linear shift in the peak of the spatial spectrum. A mathematical model is developed to quantify the relationship between the tilt angles ($\theta$, $\phi$) and the resultant spectral displacement ($k_x^{\text{peak}}$, $k_y^{\text{peak}}$). This model enables DOA estimation by locating the spectral peak, forming the foundation of the OSB-DOA algorithm \cite{ref19}.

The complex amplitude of the resonant beam defined on a reference $xy$-plane is $E(x,y) \in \mathbb{C}$, with its two-dimensional Fourier transform given by:
\begin{equation}
A_d(f_x,f_y) = \iint_{\mathbb{R}^2} E(x,y) e^{-\mathrm{j}2\pi(f_x x + f_y y)} \,\mathrm{d}x\,\mathrm{d}y,
\end{equation}
where $(f_x,f_y)$ are spatial frequencies. To facilitate the analysis of wave propagation, we introduce the physical wavenumber (or angular spatial frequency) $k=2\pi/\lambda$ and the transverse wavevector components are $(k_x, k_y)$.

A stable resonator converges to the fundamental mode. In the absence of tilt, the fundamental mode corresponds to the zeroth-order Hermite-Gaussian beam, expressed as \cite{ref25}:
\begin{equation}
E_{00}(x,y) = E_0 \exp\left[-a(x^2 + y^2)\right],
\end{equation}
where $E_0$ is the amplitude, and $a = 1/w_0^2$ determines the {\color{blue}beam waist radius} $w_0$. Applying the Fourier transform yields the spectral distribution:
\begin{equation}
A_{00}(f_x, f_y) = E_0 \frac{\pi}{a} \exp\left[-\frac{\pi^2}{a}(f_x^2 + f_y^2)\right].
\end{equation}
Equation (12) represents a Gaussian function centered at the origin $(f_x, f_y) = (0, 0)$. This indicates that for an aligned resonator, the spectral peak is located at the center of the frequency domain.

Consider the scenario where the resonator is tilted relative to the optical axis, {\color{blue}so that the beam propagation direction deviates from the $z$-axis.} Let the tilt be described by an azimuth angle $\theta$ (rotation about the $z$-axis) and an elevation angle $\phi$ (tilt relative to the $z$-axis). In the local coordinate system $(x', y', z')$ attached to the tilted resonator, the resonant beam remains the fundamental mode propagating along the $z'$-axis. Consequently, its wavevector in the local frame is given by $\mathbf{k}' = [0, 0, k]^T$, corresponding to a spectral peak at $(k'_x, k'_y) = (0,0)$.

To find the spectral shift in the global observation frame $(x,y,z)$, we apply a coordinate transformation. The global wavevector $\mathbf{k}$ is related to the local wavevector $\mathbf{k}'$ by the rotation matrix $\mathbf{R}(\theta, \phi)$:
\begin{equation}
\mathbf{k} = \begin{bmatrix} k_x \\ k_y \\ k_z \end{bmatrix} = \mathbf{R}(\theta, \phi) \mathbf{k}' = \mathbf{R}_z(\theta) \mathbf{R}_y(\phi) \begin{bmatrix} 0 \\ 0 \\ k \end{bmatrix},
\end{equation}
where $\mathbf{R}_z(\theta)$ and $\mathbf{R}_y(\phi)$ represent the rotation matrices for azimuth and elevation, respectively:
\begin{equation}
\begin{aligned}
\mathbf{R}_z(\theta) &= \begin{bmatrix} \cos\theta & -\sin\theta & 0 \\ \sin\theta & \cos\theta & 0 \\ 0 & 0 & 1 \end{bmatrix}, \\
\mathbf{R}_y(\phi) &= \begin{bmatrix} \cos\phi & 0 & \sin\phi \\ 0 & 1 & 0 \\ -\sin\phi & 0 & \cos\phi \end{bmatrix}.
\end{aligned}
\end{equation}
Substituting $\mathbf{k}' = [0, 0, k]^T$ into (13) yields the wavevector components in the global frame:
\begin{equation}
\begin{aligned}
k_x^{\text{peak}} &= k \sin\phi \cos\theta, \\
k_y^{\text{peak}} &= k \sin\phi \sin\theta.
\end{aligned}
\end{equation}
Equation (15) explicitly quantifies the linear shift of the spectral peak induced by the resonator tilt. The peak moves from the origin to a position determined by the DOA angles. Conversely, by detecting the coordinates $(k_x^{\text{peak}}, k_y^{\text{peak}})$ of the maximum amplitude in the spatial spectrum, the DOA angles can be estimated as:
\begin{equation}
\phi = \arcsin\left(\frac{1}{k}\sqrt{(k_x^{\text{peak}})^2 + (k_y^{\text{peak}})^2}\right),
\end{equation}
\begin{equation}
\theta = \arctan\left(\frac{k_y^{\text{peak}}}{k_x^{\text{peak}}}\right).
\end{equation}
Therefore, the OSB-DOA algorithm achieves high-resolution DOA estimation by mapping the spectral peak shift directly to the incident angles, bypassing the beam-width limitations in the spatial domain.

While the derivation above focuses on a single resonant beam, the framework naturally extends to multi-target scenarios based on the superposition principle of optical fields. For a system with $K$ targets, the total complex optical field $E_{\text{total}}(x,y)$ on the reference plane is the linear superposition of the individual fields $E_i(x,y)$ established by each target. Consequently, due to the linearity of the Fourier transform, the total spatial frequency spectrum $A_{\text{total}}(f_x, f_y)$ is simply the summation of the individual spectra, expressed as $A_{\text{total}}(f_x, f_y) = \sum_{i=1}^{K} A_i(f_x, f_y)$. Since each target generates a spectral peak centered at a unique spatial frequency coordinate determined by its incident angles, the multi-target DOA estimation problem effectively transforms into identifying $K$ distinct local maxima in the aggregate spectrum, allowing for simultaneous resolution of multiple targets without mutual interference.}

\subsection{Spatial Frequency Resolution and Aliasing Limits}
Based on the mathematical foundation of the discrete Fourier transform (DFT), the relationships among key parameters in the spatial frequency domain can be formally established. Let the physical size of the spatial observation window be denoted as $r_c$. The spatial frequency resolution $\Delta F${\color{blue},} defined as the smallest distinguishable interval in the frequency domain, is fundamentally constrained by the extent of the observation window and satisfies \cite{ref19}:

\begin{equation}
\Delta F = \frac{1}{r_c}
\end{equation}

{\color{blue}Here, $r_c$ refers to the length of the wavefront sensor.} {\color{blue}This resolution directly governs the distinguishability of adjacent spatial frequency components and is fixed once the observation window is specified.} On the other hand, the physical size of the observation window is determined by the product of the spatial sampling interval $\Delta$ and the number of sampling points $N$:

\begin{equation}
r_c = N \cdot \Delta
\end{equation}

Substituting (19) into (18) yields an equivalent expression for the spatial frequency resolution:

\begin{equation}
\Delta F = \frac{1}{N \cdot \Delta}
\end{equation}

Under this framework, the physical spatial frequency corresponding to the $m$-th DFT output bin can be derived from the Nyquist–Shannon sampling theorem. The theorem indicates that the maximum unambiguous spatial frequency bandwidth representable by the DFT is $\left[-\tfrac{1}{2\Delta},\, \tfrac{1}{2\Delta}\right]$, and this total bandwidth is uniformly divided into $N$ discrete bins. Therefore, the physical spatial frequency $f_{m}$ represented by the $m$-th DFT bin is given by the product of the frequency resolution and the bin index:

\begin{equation}
f_{m} = m \cdot \Delta F = m \cdot \frac{1}{N \cdot \Delta}
\end{equation}

Equation (21) provides a key mapping between the integer index $m$ in the DFT and the corresponding continuous physical spatial frequency.

By substituting this discrete mapping into the wavevector relationships established in (16) and (17), the incident angles can be explicitly computed from the discrete spectrum. Let $m_x$ and $m_y$ denote the indices of the spectral peak along the $f_x$ and $f_y$ axes, respectively, relative to the zero-frequency center (DC component). The physical spatial frequencies are given by $f_x = m_x \Delta F$ and $f_y = m_y \Delta F$. Consequently, the practical calculation formulas for the OSB-DOA algorithm are derived as:
\begin{equation}
\phi = \arcsin\left( \lambda \cdot \Delta F \cdot \sqrt{m_x^2 + m_y^2} \right),
\end{equation}
\begin{equation}
\theta = \arctan\left(\frac{m_y}{m_x}\right),
\end{equation}
where $\lambda$ is the wavelength. These equations constitute the discrete implementation of the theoretical model derived in Section III-B. {\color{blue}In summary, the OSB-DOA algorithm operates by first performing a two-dimensional DFT on the sampled optical field, identifying the peak indices $(m_x, m_y)$ in the spatial frequency spectrum, and finally applying Equations (22) and (23) to compute the estimated angles. The complete procedure is formalized in Algorithm~1, which also annotates the computational complexity of each step.}

{\color{blue}
\begin{algorithm}[t]
\caption{Optical Spectrum-Based DOA Estimation}
\label{alg:osb_doa}
\begin{algorithmic}[1]
\STATE \textbf{Input:} Sampled optical field $E[n_x, n_y]$ of size $N \times N$, wavelength $\lambda$, sampling interval $\Delta$.
\STATE \textbf{Output:} Estimated Elevation $\hat{\phi}$ and Azimuth $\hat{\theta}$.
\STATE \textit{// Step 1: Spectral Transformation (Complexity: $\mathcal{O}(N^2 \log N)$)}
\STATE Compute 2D FFT: $\mathbf{A} = \text{FFT2}(E)$
\STATE Shift zero-frequency component to center: $\mathbf{A}_{\text{shift}} = \text{fftshift}(\mathbf{A})$
\STATE \textit{// Step 2: Peak Detection (Complexity: $\mathcal{O}(N^2)$)}
\STATE Find the global maximum in amplitude spectrum $|\mathbf{A}_{\text{shift}}|$
\STATE Get peak indices $(m_x, m_y)$ relative to the DC center $(0,0)$
\STATE \textit{// Step 3: DOA Calculation (Complexity: $\mathcal{O}(1)$)}
\STATE Calculate spatial frequency resolution: $\Delta F = 1 / (N \cdot \Delta)$
\STATE Estimate Elevation: $\hat{\phi} = \arcsin(\lambda \cdot \Delta F \cdot \sqrt{m_x^2 + m_y^2})$
\STATE Estimate Azimuth: $\hat{\theta} = \operatorname{atan2}(m_y, m_x)$
\RETURN $\hat{\phi}, \hat{\theta}$
\end{algorithmic}
\end{algorithm}
}

{\color{blue}Moreover, the Nyquist-Shannon sampling theorem also implies that the maximum spatial frequency $f_{\text{max}}$ that can be accurately represented without aliasing is entirely determined by the spatial sampling interval $\Delta$:

\begin{equation}
f_{\text{max}} = \frac{1}{2\,\Delta}
\end{equation}

Reducing $\Delta$ increases the Nyquist limit, expanding the alias-free bandwidth.} {\color{blue}For a two-dimensional resonant beam with an elevation angle $\phi$ and an azimuth angle $\theta$, the spatial frequency components projected onto the orthogonal axes are $f_x = \sin\phi \cos\theta / \lambda$ and $f_y = \sin\phi \sin\theta / \lambda$. To avoid aliasing, the sampling condition must be satisfied along both axes simultaneously, i.e., $\max(|f_x|, |f_y|) < f_{\text{max}}$. Consequently, the system's anti-aliasing capability is dictated by the dominant projected component, imposing a constraint on the sampling interval:

\begin{equation}
\Delta < \frac{1}{2 \max(|f_x|, |f_y|)}
\end{equation}

This property implies that the aliasing threshold is azimuth-dependent. For instance, oblique incidence reduces the projected frequency components compared to the worst-case radial alignment, effectively relaxing the sampling requirement. Under a fixed resolution $\Delta F$, reducing $\Delta$ ensures that these high-frequency components remain within the unaliased spectral range.

Beyond the aliasing limit, it is essential to quantify the angular resolution sensitivity. This sensitivity measures how the finite spatial frequency resolution $\Delta F$ translates into angular uncertainty. To perform this analysis, we first map the wavevector relationships derived in (16) and (17) into the spatial frequency domain using the relations $k=2\pi/\lambda$ and $k_{x,y}=2\pi f_{x,y}$.

For the elevation angle $\phi$, the wavevector magnitude relationship in (16) transforms to:
\begin{equation}
\sin\phi = \frac{1}{k}\sqrt{(k_x^{\text{peak}})^2 + (k_y^{\text{peak}})^2} = \lambda \sqrt{f_x^2 + f_y^2}.
\end{equation}
Applying first-order error propagation with respect to the spatial frequency magnitude yields the resolution limit:
\begin{equation}
\Delta \phi \approx \left| \frac{\partial \phi}{\partial \sqrt{f_x^2 + f_y^2}} \right| \Delta F = \frac{\lambda}{\cos\phi} \Delta F = \frac{\lambda}{r_c \cos\phi}.
\end{equation}

For the azimuth angle $\theta$, Eq. (17) indicates that $\theta = \arctan(f_y/f_x)$. Geometrically, this corresponds to the phase angle in the spatial frequency plane. At the spectral radius determined by $\phi$ (where $f_r = \sin\phi/\lambda$), the angular uncertainty is constrained by the tangential resolution. Specifically, the arc length corresponding to $\Delta \theta$ at this radius is limited by the frequency resolution $\Delta F$:
\begin{equation}
\begin{aligned}
&\Delta \theta \cdot \frac{\sin\phi}{\lambda} \approx \Delta F \\
\Rightarrow \quad &\Delta \theta \approx \frac{\Delta F}{(\sin\phi)/\lambda} = \frac{\lambda}{r_c \sin\phi}.
\end{aligned}
\end{equation}

Combined with the derived angular resolution bounds, this framework provides a direct and quantifiable theoretical basis for sensor design and parameter configuration in OSB-DOA.}

\subsection{{\color{blue}TM Direction Correction}}
To establish an accurate resonant beam propagation model in the RB-HWDOA system, the directional distortion induced by the convex lens in the retroreflector must be considered. To address this issue, this paper introduces a TM module, which performs direction correction before the beam enters the sensing module. We define the convex lens inside the retroreflector as $L_{11}$, and the convex lenses in the TM architecture as $L_{12}$, $L_{21}$, and $L_{22}$, as shown in Fig.~7. {\color{blue}Here, $f_1$ denotes the focal length of lenses $L_{11}$ and $L_{12}$, and $f_2$ denotes the focal length of lenses $L_{21}$ and $L_{22}$. The four lenses are arranged with inter-lens spacings of $2f_1$, $f_1+f_2$, and $2f_2$, respectively.} The entire analysis is based on the matrix optics framework for off-axis beams: the beam state is described by the column vector $\mathbf{r}=(x,\eta)^\mathrm{T}$, where $x$ is the lateral displacement from the optical axis and $\eta$ is the incident angle relative to the optical axis (in radians). The input-output relationship of any optical subsystem can be expressed as $\mathbf{r}_2 = \mathbf{M} \mathbf{r}_1$, where $\mathbf{M} \in \mathbb{R}^{2\times2}$ is the {\color{blue}ray transfer} matrix. {\color{blue}Each optical element is represented by its own $2\times2$ matrix, and the total system matrix is obtained by multiplying the individual matrices in propagation order.}

\begin{figure}[!t]
\centering
\includegraphics[width=3in]{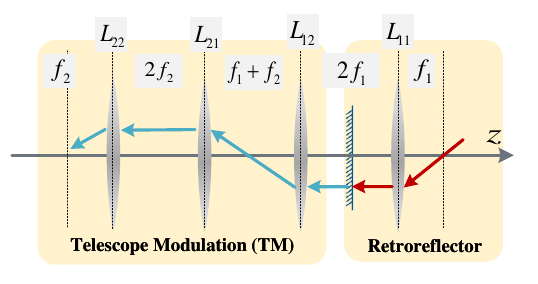}
\caption{{\color{blue}Detailed optical path of the Tx unit. The retroreflector lens $L_{11}$ (focal length $f_1$) and three TM lenses $L_{12}$ (focal length $f_1$), $L_{21}$ (focal length $f_2$), and $L_{22}$ (focal length $f_2$) are arranged along the optical axis $z$ with inter-lens spacings of $2f_1$, $f_1+f_2$, and $2f_2$, respectively, satisfying the afocal condition. The end-face mirror functions as a partially reflective output coupler. The red arrows represent the intracavity resonant beam sustained by reflection, while the blue arrows denote the sensing beam transmitted through the output coupler to the TM module.}}
\label{fig_rbfd}
\end{figure}

For a thin convex lens with focal length $f$, the front and back focal planes are taken as the incident and outgoing reference planes, respectively. The transfer matrix for this lens is:
\begin{equation}
    \begin{aligned}
        \mathbf{M}(f) &=
        \begin{bmatrix}
            1 & f \\
            0 & 1
        \end{bmatrix}
        \begin{bmatrix}
            1 & 0 \\
            -\frac{1}{f} & 1
        \end{bmatrix}
        \begin{bmatrix}
            1 & f \\
            0 & 1
        \end{bmatrix} \\
        &=
        \begin{bmatrix}
            0 & f \\
            -\frac{1}{f} & 0
        \end{bmatrix}.
    \end{aligned}
\end{equation}

{\color{blue}Equation (29) shows that a single lens swaps the beam's position and angle, scaled by $f$. In the retroreflector, the beam passes through $L_{11}$ (focal length $f_1$), propagates to the end-face plane mirror, and returns. This mirror also serves as an output coupler (reflectivity $R_1$, transmittance $1-R_1$); for the direction-correction derivation it is modeled as the identity matrix $\mathbf{I}$, while its amplitude effect is treated in Section~III-E. The single-pass matrix is $\mathbf{M}_{L_{11}} = \mathbf{M}(f_1)$, and the round-trip matrix is:}

\begin{equation}
    \begin{aligned}
        \mathbf{M}_r &= \mathbf{M}_{L_{11}} \mathbf{I} \mathbf{M}_{L_{11}} \\
        &= \begin{bmatrix}
            0 & f_1 \\
            -\frac{1}{f_1} & 0
        \end{bmatrix}
        \begin{bmatrix}
            0 & f_1 \\
            -\frac{1}{f_1} & 0
        \end{bmatrix} \\
        &= \begin{bmatrix}
            -1 & 0 \\
            0 & -1
        \end{bmatrix}.
    \end{aligned}
\end{equation}

From equation (30), it is clear that the retroreflector inverts the beam state, i.e., $\mathbf{r}_2 = -\mathbf{r}_1$, thus completing the direction reversal. {\color{blue}For the sensing path, the beam passes through $L_{11}$ once and is transmitted through the output coupler into the TM module. The effective single-pass matrix is:}
\begin{equation}
    \begin{aligned}
        \mathbf{M}_l &= \mathbf{M}(f_1) \\
        &= \begin{bmatrix}
            0      & f_1 \\
            -\frac{1}{f_1} & 0
        \end{bmatrix}.
    \end{aligned}
\end{equation}
Clearly, $\mathbf{M}_l$ {\color{blue}converts the incident angle $\eta_1$ into a lateral displacement $x_2 = f_1 \eta_1$, while the original position $x_1$ is converted into an exit angle $\eta_2 = -x_1/f_1$. This position--angle interchange is the fundamental cause of the directional distortion introduced by the retroreflector.}

{\color{blue}
The TM module ($L_{12}$, $L_{21}$, $L_{22}$) has the overall matrix:
\begin{equation}
    \begin{aligned}
        \mathbf{M}_{\text{TM}}
        &= \mathbf{M}(f_2) \mathbf{M}(f_2) \mathbf{M}(f_1) \\
        &= \begin{bmatrix}
            0 & f_2 \\
            -\frac{1}{f_2} & 0
        \end{bmatrix}
        \begin{bmatrix}
            0 & f_2 \\
            -\frac{1}{f_2} & 0
        \end{bmatrix}
        \begin{bmatrix}
            0 & f_1 \\
            -\frac{1}{f_1} & 0
        \end{bmatrix} \\
        &= \begin{bmatrix}
            0 & -f_1 \\
            \frac{1}{f_1} & 0
        \end{bmatrix}.
    \end{aligned}
\end{equation}
The product of two $\mathbf{M}(f_2)$ matrices gives the negative identity matrix, and multiplying this with $\mathbf{M}(f_1)$ yields equation (32). The total transmission matrix for the RB passing through $L_{11}$ and the TM module is:
\begin{equation}
    \begin{aligned}
        \mathbf{M}_{\text{total}}
        &= \mathbf{M}_{\text{TM}} \mathbf{M}_l \\
        &= \begin{bmatrix}
            0 & -f_1 \\
            \frac{1}{f_1} & 0
        \end{bmatrix}
        \begin{bmatrix}
            0 & f_1 \\
            -\frac{1}{f_1} & 0
        \end{bmatrix} \\
        &= \begin{bmatrix}
            1 & 0 \\
            0 & 1
        \end{bmatrix}.
    \end{aligned}
\end{equation}

Substituting the initial state $\mathbf{r}_1 = (x_1, \eta_1)^\mathrm{T}$ into equation (33), we obtain:
\begin{equation}
    \begin{aligned}
        \mathbf{r}_2 &= \mathbf{M}_{\text{total}} \mathbf{r}_1 \\
        &= \begin{bmatrix}
            x_1 \\
            \eta_1
        \end{bmatrix}.
    \end{aligned}
\end{equation}
Equation (34) shows that the TM module preserves the original state: $x_2 = x_1$ and $\eta_2 = \eta_1$. In other words, the beam direction is accurately restored. If the RB passes through the front focal plane of $L_{11}$ during incidence, then $x_1 = 0$, and consequently, $x_2 = 0$. The beam will focus without lateral displacement at the back focal plane of $L_{22}$, and the RBs emitted from all Tx will naturally converge at this plane, providing an ideal sensing plane for DOA estimation. 
}

{\color{blue}
Although the derivation in Eqs. (33)–(34) is presented for a single optical axis, the spherical deployment of Tx units ensures that the optical axis of each Tx intersects at the center of the sensing module. This shared intersection point naturally defines the origin of the global DOA coordinate system: the elevation angle $\phi$ and azimuth angle $\theta$ estimated by the OSB-DOA algorithm are measured with respect to this common origin, regardless of which Tx unit carries the resonant beam. Because the TM module preserves the ray state ($\eta_2 = \eta_1$, $x_2 = x_1$), each corrected beam is refocused to its local back focal plane, which coincides with the common sensing plane. Consequently, the fields from all Tx units are superposed directly at the wavefront sensor in a unified angular reference frame, and the OSB-DOA algorithm operates on the composite field without requiring per-Tx coordinate transformations or data fusion.}

\subsection{Noise Analysis}
In this section, we present the output power model of the resonant beam and quantify the noise power of the CMOS-based wavefront sensor to derive the SNR model of RB-HWDOA.

The SNR is a critical factor that impacts the accuracy of the RB-HWDOA. Let $S$ represent the signal power on {\color{blue}the sensing module} and {\color{blue}$P_N$} represent the power of the noise. The value of $S$ can be calculated from the laser output power of the resonator, expressed {\color{blue}as}:
\begin{equation}
S = \rho P_{\text{out}},
\end{equation}
where $\rho$ is the {\color{blue}attenuation factor}, and ${\color{blue}P_{\text{out}}}$ is the {\color{blue}output} power of {\color{blue}the} resonant cavity, {\color{blue}expressed as follows} \cite{ref18}:
\begin{equation}
\begin{aligned}
P_{\text{out}} = &\frac{A_b I_s (1 - R_1) \sqrt{V_r}}{1 - R V_r + \sqrt{R V_r} \left[ 1/(V_r V_s)-V_s \right]} \\
                 &\cdot \left[ \frac{\eta_{\text{excit}} P_{\text{in}}}{A_g I_s} - \ln \left| \sqrt{R V_{s}^{2} V_r} \right| \right],
\end{aligned}
\end{equation}
{\color{blue}where $A_{b}$ is the resonant beam area on the gain medium, $A_{g}$ is the cross-sectional area of the gain medium, $I_{s}$ is the saturation intensity of the gain medium, $\eta_{\text{excit}}$ represents the excitation efficiency at the gain medium, $V_s$ denotes the single-pass loss factor of the gain medium, and $V_{r}$ is the over-the-air transmission efficiency characterizing the round-trip propagation loss between the Tx and Rx. The mirror reflectivity parameters are defined as follows: $R_1$ is the reflectivity of the output coupler, $R_2$ is the reflectivity of the high reflector, and $R = R_1 R_2$ represents the effective round-trip mirror reflectivity.}

After accounting for the signal power, we proceed to model the noise power contributed by the CMOS-based wavefront sensor. The principal noise sources in a typical CMOS image sensor comprise photon shot noise, dark current noise, read noise, and thermal noise. The overall noise power can be summarized as follows \cite{ref18,ref19}:
\begin{equation}
{\color{blue}P_N} = (q \eta_{\text{cmos}} \Phi + i_d) T_{\text{int}} + \sigma_{\text{read}}^2 + \frac{4k_B T B}{R_{\text{load}}}
\end{equation}
where \(q\) represents the electron charge, \(\eta_{\text{cmos}}\) is the quantum efficiency of the CMOS sensor, \(\Phi\) denotes the incident photon flux, \(i_d\) is the dark current, \(T_{\text{int}}\) is the integration time, \(\sigma_{\text{read}}^2\) is the read noise variance, \(k_B\) is Boltzmann’s constant, \(T\) is the absolute temperature, \(R_{\text{load}}\) is the load resistance, and \(B\) is the system bandwidth. This model captures the major noise mechanisms in CMOS-based optical sensing.

The SNR of the RB-HWDOA system is given by:
\begin{equation}
\text{SNR} = \frac{\rho P_{\text{out}}}{{\color{blue}P_N}},
\end{equation}
Here, $P_{\text{out}}$ denotes the output power of the resonant cavity, and {\color{blue}$\rho$ represents the attenuation factor.}

{\color{blue}The OSB-DOA algorithm estimates DOA from spectral-peak positions rather than amplitudes (cf.\ Section~III-B). Therefore, signal power variations introduced by the output-coupler transmittance $(1-R_1)$, the optical-path attenuation factor $\rho$, and the angle-dependent transmission efficiency $V_r$ in Eqs.~(35)--(36) affect only the SNR at the wavefront sensor, not the estimated DOA angles. In addition, resonant beams are inherently energy-concentrated: the resonant cavity suppresses scattered radiation and confines energy to a highly collimated beam, yielding favorable SNR under normal operating conditions. This robustness is demonstrated by the multi-SNR simulation results in Section~IV-B.}

{\color{blue}
\subsection{High-Resolution Mechanism of the OSB-DOA Algorithm}

The fundamental transverse mode of a resonator can be accurately modeled by a Gaussian beam:
\begin{equation}
E_{00}(x, y) = E_{0} \exp\left[-\frac{x^2 + y^2}{w_0^2}\right],
\end{equation}
{\color{blue}where $E_0$ is the peak amplitude, and $w_0$ is the beam waist radius. Since the far-field divergence angle $\beta \propto 1/w_0$, a highly collimated resonant beam has a large $w_0$. In the OSCB-DOA algorithm, the large $w_0$ of beams from adjacent DOAs causes significant spatial overlap, limiting the angular resolution.}

From the perspective of Fourier optics, the spatial spectrum is obtained by performing a Fourier transform on the spatial profile. Consistent with the definition in Eq. (11), the frequency-domain representation is:
\begin{equation}
A_{00}(f_x, f_y) = E_0 \pi w_0^2 \exp\left[- \pi^2 w_0^2(f_x^2 + f_y^2)\right],
\end{equation}
where $(f_x, f_y)$ are the spatial frequencies. The spectral width $\Delta f$ of this distribution is proportional to $1/w_0$. This indicates that a wider beam in the spatial domain (large $w_0$) corresponds to a narrower and more localized distribution in the spatial frequency domain.

Since the physical direction is mapped to the spatial frequency as $k_x = 2\pi f_x = k \sin\phi \cos\theta$ (as derived in Eq. (15)), the spectral width $\Delta f$ directly determines the angular resolution limit. By leveraging the mathematical relationship $\Delta f \propto 1/w_0$, the OSB-DOA algorithm turns the disadvantage of a large spatial footprint into a significant advantage. The large spatial beam waist is transformed into highly sharpened spectral features. This process effectively maps overlapping spatial signals into distinct and easily separable spectral peaks, thereby significantly improving the minimum resolvable DOA performance.
}

\section{Numerical Analysis and Results}
We begin by verifying the feasibility of the OSB-DOA technique, followed by an investigation into the impact of key factors such as spatial resolution and SNR on the accuracy of {\color{blue}DOA} estimation. Next, we evaluate the computational complexity of the OSB-DOA method by comparing its floating-point operations (FLOPs) requirement to those of other {\color{blue}DOA} estimation methods. Finally, we discuss the performance of the OSB-DOA algorithm in multi-target scenarios, focusing on its ability to distinguish sources with closely spaced {\color{blue}angles} and its robustness to noise and resolution limitations.

\begin{table}[htbp]
\centering
\caption{{\color{blue}PARAMETERS FOR RBS SIMULATION} \cite{ref17,ref18,ref19}}
\begin{tabular}{@{}lll@{}}
\toprule
\textbf{Symbol} & \textbf{Parameter} & \textbf{Value} \\ \midrule
\( r_m \)       & Cat's eye radius   & 4.5 mm            \\
\( r_l \)       & lens radius   & 4.5 mm            \\
\( r_c \)       & Wavefront sensor length   & 4.5 mm            \\
\( r_g \)       & Gain medium radius & 4.5 mm            \\
\( f_1 \)       & Focal length of $L_{11}$/$L_{12}$ & 15 mm           \\
\( f_2 \)       & Focal length of $L_{21}$/$L_{22}$ & 150 mm          \\
\( \lambda \)   & Resonant beam wavelength & 1064 nm       \\
\( R_1 \)       & Reflectivity of output coupler  & 0.95            \\
\( R_2 \)       & Reflectivity of high reflector & 0.99            \\
\( \eta_{excit} \) & Excitation efficiency & 0.72       \\
\( V_s \)       & Loss factor in medium & 0.99           \\
\( I_s \)       & Medium saturated intensity & \( 1.26 \times 10^6 \) \( \frac{W}{m^2} \) \\
\( \rho \)      & Attenuation factor & \( 10^{-4} \)    \\
\( P_{\text{in}} \) & Pump power & 100 W           \\
\( A_b \)       & Beam area on gain medium & \makecell[l]{\( \pi r_l^2 \)} \\
\( A_g \)       & Gain medium area & \( \pi r_g^2 \) \\
\( N_s \)       & Sense module pixel number & 16384/8192/4096              \\
\( S_N \)       & FFT sampling number & 16384/8192/4096            \\
\( G \)         & Computation window expand factor & 3    \\ \bottomrule
\end{tabular}
\end{table}

\subsection{Parameters}
The RB-HWDOA system is designed based on the RBS architecture, utilizing the resonant beam for information transmission. Experimental results verify that the resonant beam can be spontaneously established when the Tx and Rx are misaligned and can automatically align thereafter at 2 m, enabling {\color{blue}DOA estimation}. Additionally, simulations confirmed that the resonant beam can achieve emission and energy transfer at 10 m \cite{ref18,ref19}, and the TM system can change the light phase and propagation direction based on reference \cite{ref17}. Therefore, our simulation parameters are based on reference \cite{ref17,ref18,ref19}, as shown in Table~I.

Key component parameters in the resonator, such as the gain medium, are derived from an exemplary diode-pumped Nd:~YVO$_4$ laser system used for numerical analysis. We used the Fox-Li iterative algorithm in the simulation to obtain the RBS self-reproducing mode. Following the guidance in reference \cite{ref19}, appropriate sampling and zero-padding were applied to avoid aliasing in FFT, and the sampling accuracy met the requirements for accurate resonator mode analysis.

\subsection{Analysis of {\color{blue}DOA} Accuracy}
The RB-HWDOA system employs the OSB-DOA method for direction-of-arrival estimation. Therefore, we first validate the feasibility of the OSB-DOA method, then investigate the impact of SNR and the sampling rate of the sensing module on {\color{blue}DOA} estimation. Finally, we demonstrate the advantage of the OSB-DOA method by comparing the required floating-point operations with other {\color{blue}DOA} estimation methods. The RBS pump power $P_{\text{in}}$ is fixed at 100 W.

\begin{figure}[!t]
\centering
\includegraphics[width=3in]{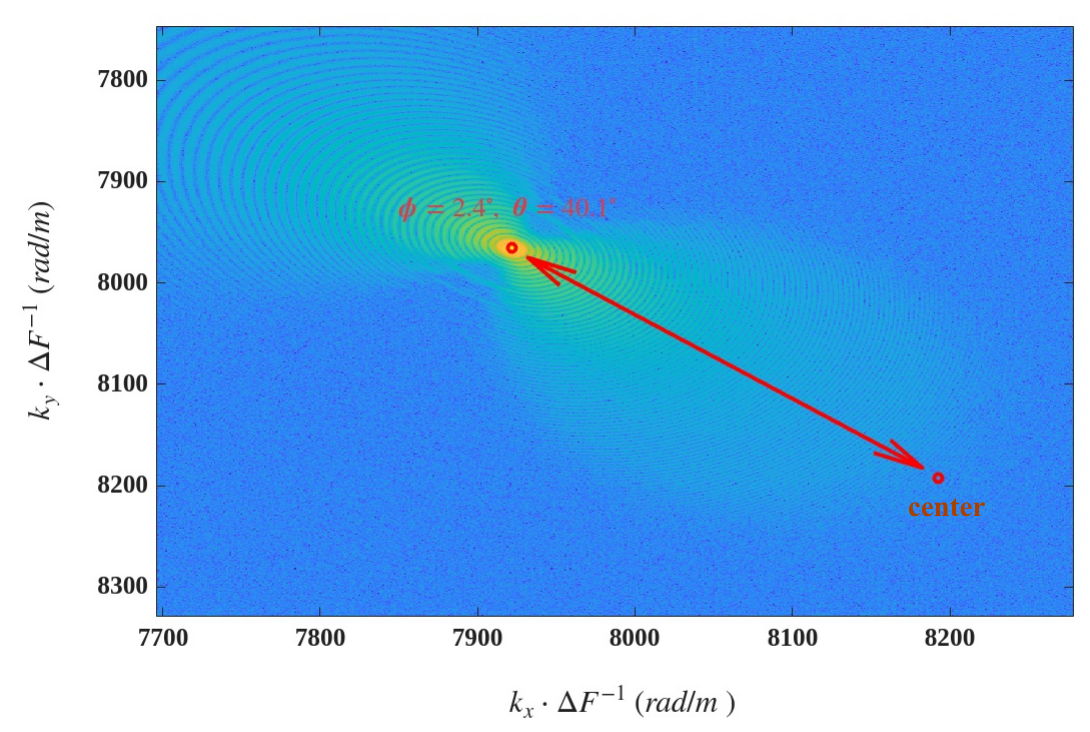}
\caption{The spatial spectrum of a single resonant beam signal, where the horizontal and vertical axes represent the sampling rates. Here, $k_x$ and $k_y$ are the angular spatial frequencies, {\color{blue}the center denotes the zero-frequency component}, and $\Delta F$ represents the spatial spectral resolution.}
\end{figure}

Fig.~8 illustrates the spatial spectrum of the resonant beam when the elevation angle $\phi = 2.4^{\circ}$ and the azimuth angle $\theta = 40^{\circ}$. The horizontal and vertical axes represent the sampled values of spatial frequencies. Clear and distinct spectral peaks are observed in the figure, with noticeable shifts from the central frequency. {\color{blue}These spectral shifts correspond to the incident elevation and azimuth angles through Eqs.~(22) and (23), confirming the feasibility of the OSB-DOA method.}

\begin{figure}[!t]
\centering
    \subfloat[$\phi$ Error]{
        \includegraphics[width=3in]{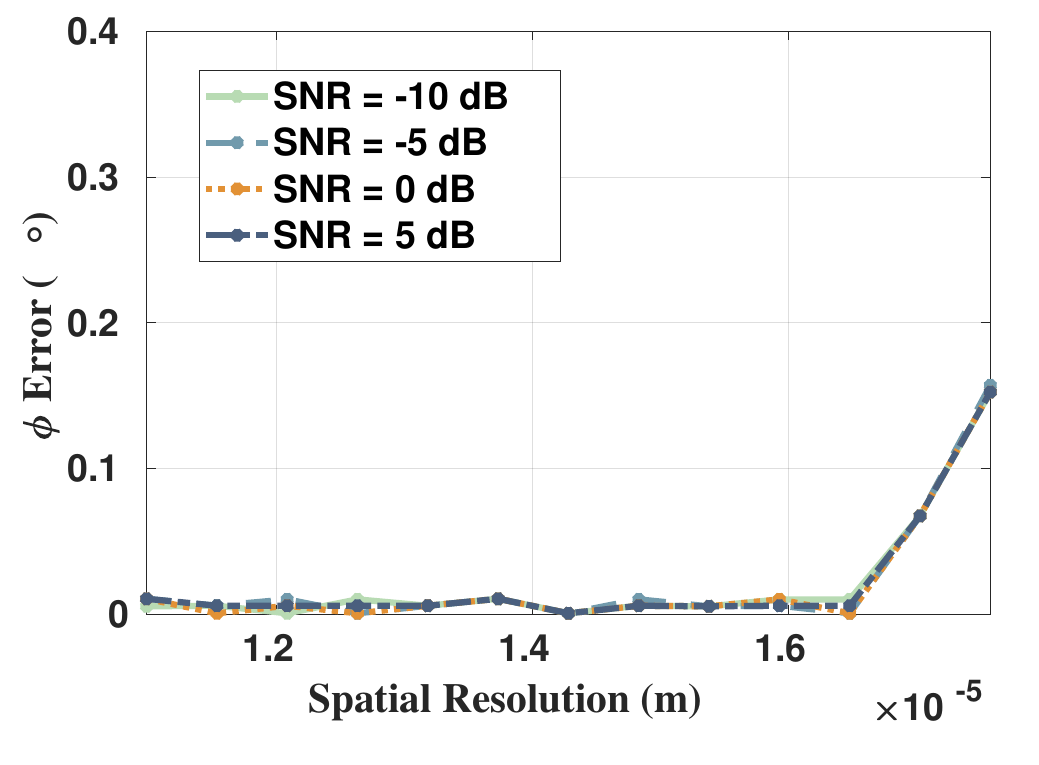}
    }\\
    \subfloat[$\theta$ Error]{
        \includegraphics[width=3in]{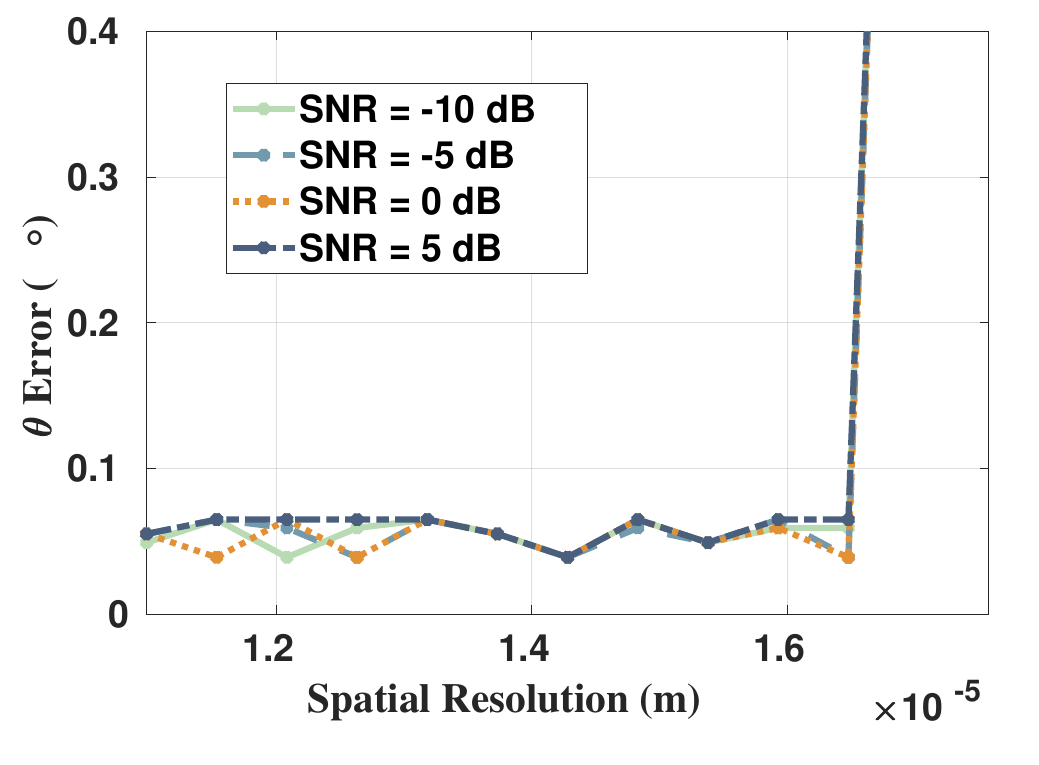}
    }
\caption{{\color{blue}Impact of spatial resolution and SNR on DOA estimation error.}}
\end{figure}

{\color{blue}The derivations in Sections~III-C show that the sampling interval primarily determines the maximum estimable $\phi$. To verify this relationship, we increase $\Delta$ by reducing the sampling density and compare the DOA estimation errors under different SNR levels, as shown in Fig.~9.} Fig.~9a and Fig.~9b show the variation of {\color{blue}elevation} and azimuth angle estimation errors with respect to SNR and sampling interval. From the figures, it can be observed that when the sampling interval is less than $2.75\times10^{-6}$~m, the estimation errors for {\color{blue}elevation} and azimuth angles stabilize at $0.01^{\circ}$ and $0.05^{\circ}$, respectively, and do not show significant changes with varying SNR. However, when the sampling interval exceeds this threshold, both types of errors increase sharply. This phenomenon verifies the impact of spectral aliasing on DOA estimation accuracy: within a certain sampling rate range, reducing the sampling rate does not significantly affect estimation performance. {\color{blue}However, when the sampling interval $\Delta$ is too large to satisfy the anti-aliasing condition $\sin\phi/\lambda < f_{\text{max}} = 1/(2\Delta)$ (cf. Eq.~(25)), aliasing effects occur, causing a significant increase in estimation errors.} Furthermore, the comparison of DOA errors under different SNR levels indicates that random Gaussian noise does not significantly shift the peak of the spatial spectrum, demonstrating that the proposed OSB-DOA method exhibits strong noise robustness. Nevertheless, when estimating {\color{blue}a} large {\color{blue}elevation} angle, it is essential to choose an appropriate sampling rate to avoid aliasing.

\begin{figure}[!t]
\centering
\includegraphics[width=3in]{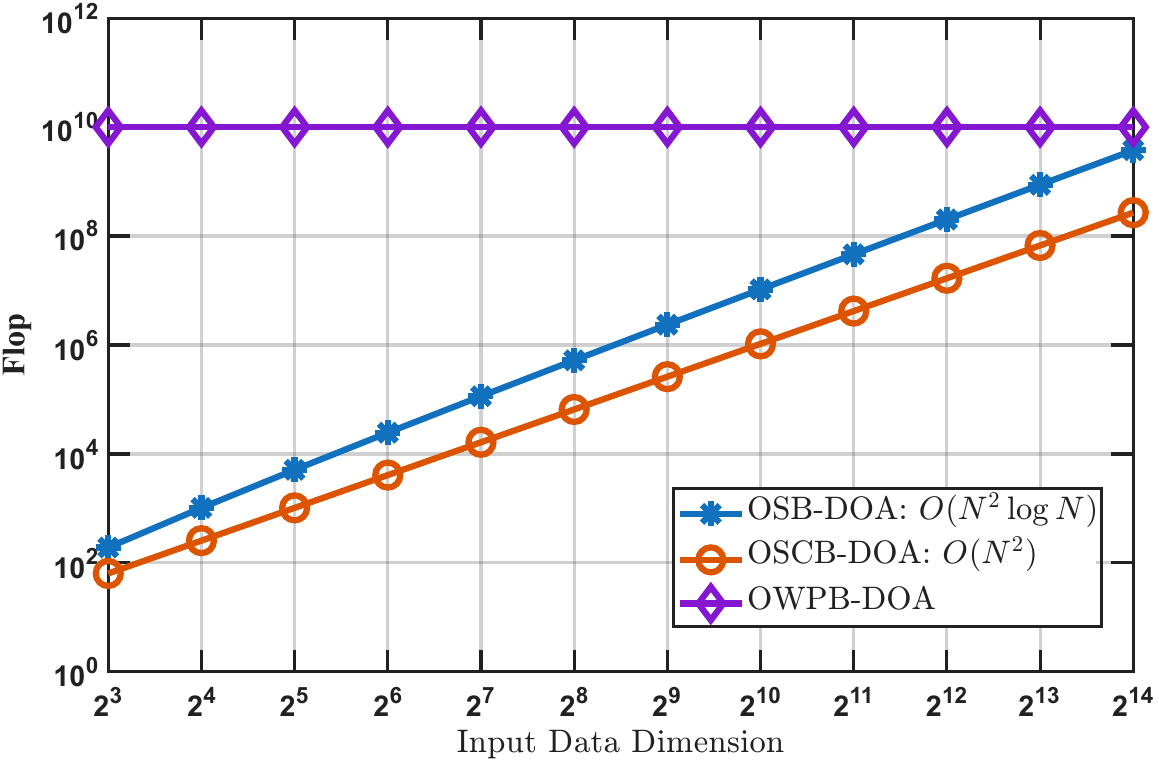}
\caption{The relationship between the OSB-DOA, OSCB-DOA, and OWPB-DOA algorithms and the number of FLOPs required and the input data dimension.}
\end{figure}

{\color{blue}
To comprehensively evaluate the proposed method and interpret the computational complexity results presented in Fig.~10, we detail the implementation mechanisms and algorithmic complexity of the proposed method alongside two established benchmarks:

\begin{enumerate}
\item \textbf{Proposed OSB-DOA (This Work):} As detailed in Algorithm~1, this method estimates DOA by performing a 2D FFT on the $N \times N$ sampled optical field followed by spectral peak detection and closed-form angular inversion. The overall computational complexity is $\mathcal{O}(N^2 \log N)$, dominated by the FFT step, ensuring high efficiency even with large input dimensions.

\item \textbf{OWPB-DOA \cite{ref17}:} This method utilizes the phase distribution for high-precision estimation. It treats the wavefront sensor pixels analogously to an antenna array and applies the subspace-based MUSIC algorithm. Since the input is a two-dimensional wavefront of $D \times D$ pixels, the algorithm vectorizes the data into a $D^{2}$-dimensional observation vector and constructs a $D^{2} \times D^{2}$ covariance matrix. The subsequent eigenvalue decomposition therefore scales as $\mathcal{O}\!\left((D^{2})^{3}\right)=\mathcal{O}(D^{6})$, resulting in considerably higher computational cost, as reflected in the steep slope in Fig.~10.

\item \textbf{OSCB-DOA \cite{ref18,ref23,ref24}:} This benchmark represents intensity-based estimation. It employs the Gray-weighted Centroid Algorithm to compute the energy center of the beam spot. This involves a weighted average calculation in a single pass over the array, resulting in the lowest complexity of $\mathcal{O}(N^2)$, which scales linearly with the total number of pixels.
\end{enumerate}

Fig.~10 quantitatively compares the required floating-point operations (FLOPs) for these methods. As analyzed above, while the OSCB-DOA requires the fewest operations due to its simple centroid calculation, the proposed OSB-DOA maintains a comparable order of magnitude with $\mathcal{O}(N^2 \log N)$ complexity. The OWPB-DOA, owing to its $\mathcal{O}(D^{6})$ complexity arising from the 2D MUSIC eigenvalue decomposition, incurs a substantially higher computational cost that increases rapidly with input dimension.}

\subsection{Multi-target DOA Estimation Analysis}

\begin{figure}[!t]
\centering
\includegraphics[width=3in]{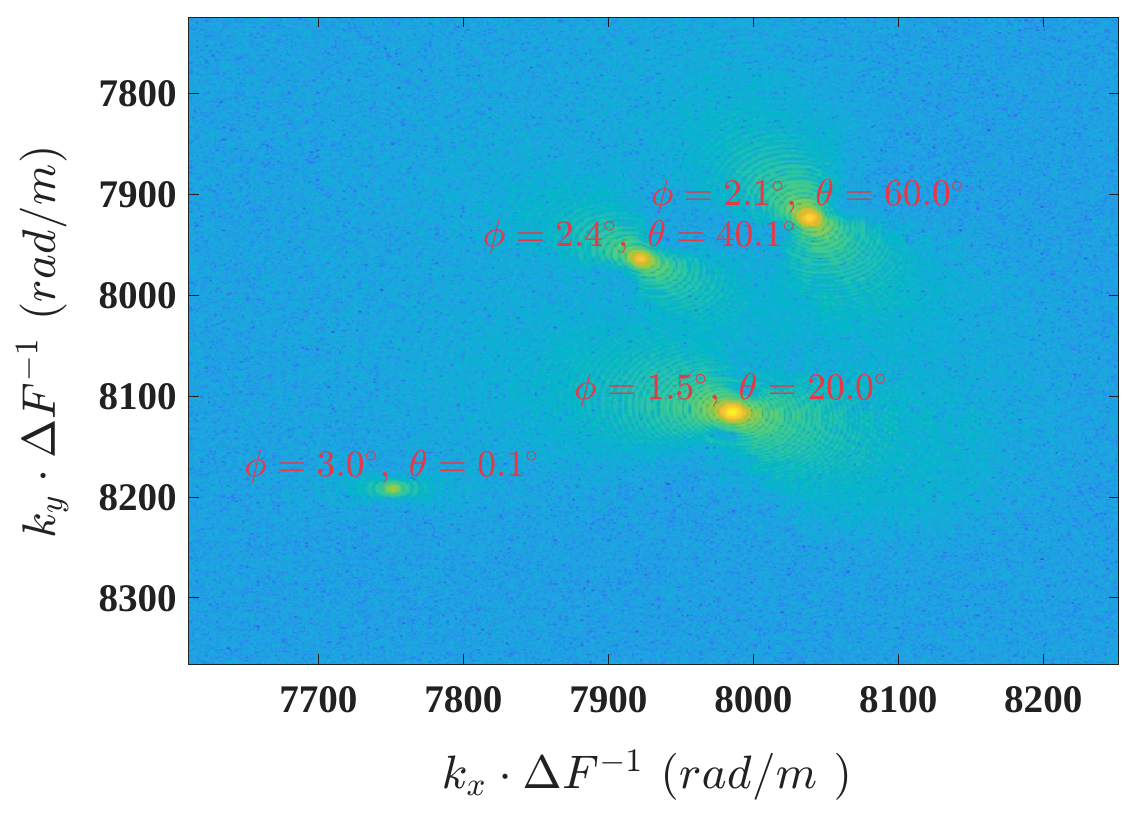}
\caption{Spatial spectrum representation of four light sources with varying azimuth and elevation angles.}
\end{figure}

The RB-HWDOA system utilizes the OSB-DOA method to perform simultaneous DOA estimation for {\color{blue}multiple targets}. Therefore, this subsection analyzes the performance of the OSB-DOA method in multi-target scenarios.

{\color{blue}As established in Section~III-B, the linearity of the Fourier transform ensures that superimposed resonant beams produce distinct spectral peaks in the frequency domain, as shown in Fig.~11.} By identifying the four dominant frequency regions in the spectrum, the associated azimuth and elevation angles can be sequentially computed using Equations {\color{blue}(22) and (23)}, thereby enabling DOA estimation in multi-target scenarios.

\begin{figure}[!t]
\centering
\includegraphics[width=3in]{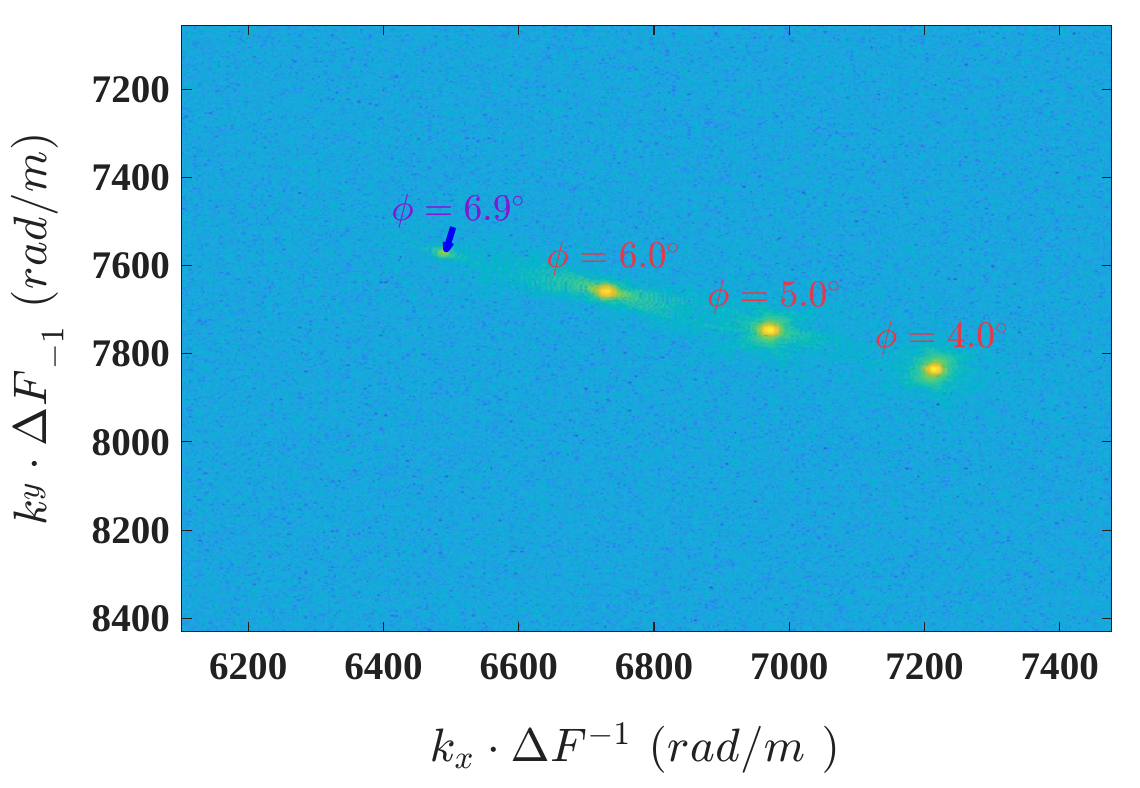}
\caption{The spatial spectrum for four light sources.}
\end{figure}

\begin{figure}[!t]
\centering
\includegraphics[width=3in]{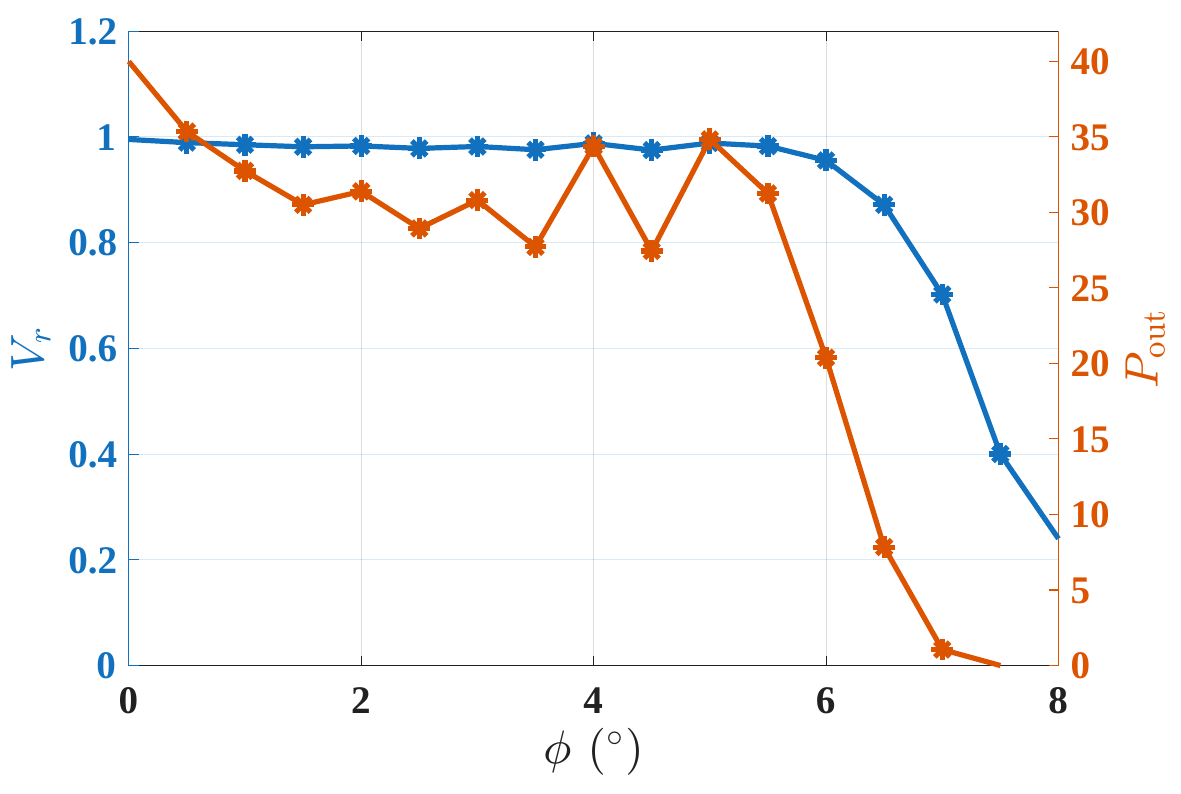}
\caption{{\color{blue}Transmission efficiency $V_r$ and output power $P_{\text{out}}$ (W) of the resonant beam as a function of the angle $\phi$.}}
\end{figure}

\begin{figure*}[!t]
    \centering
    \subfloat[OSB-DOA: $ \Delta \phi = 0.1^{\circ}$]{
        \includegraphics[width=0.26\textwidth]{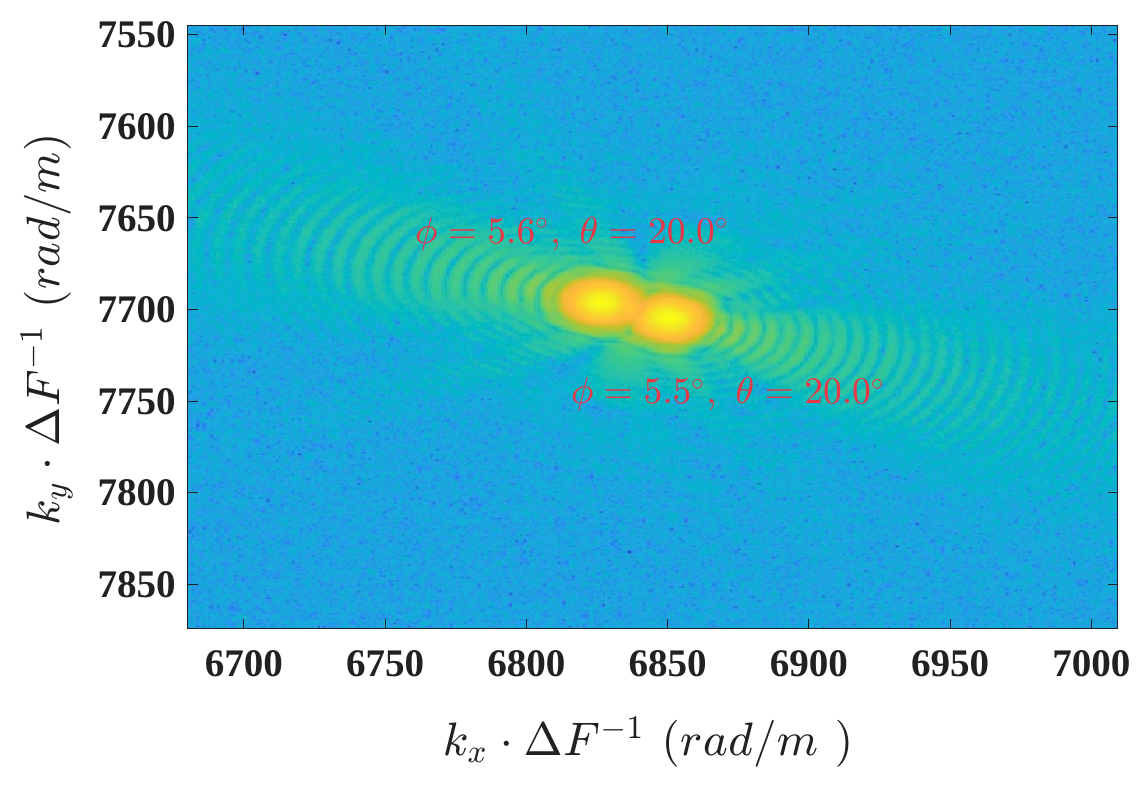}
    }
    \subfloat[OWPB-DOA: $ \Delta \phi = 0.5^{\circ}$]{
        \includegraphics[width=0.26\textwidth]{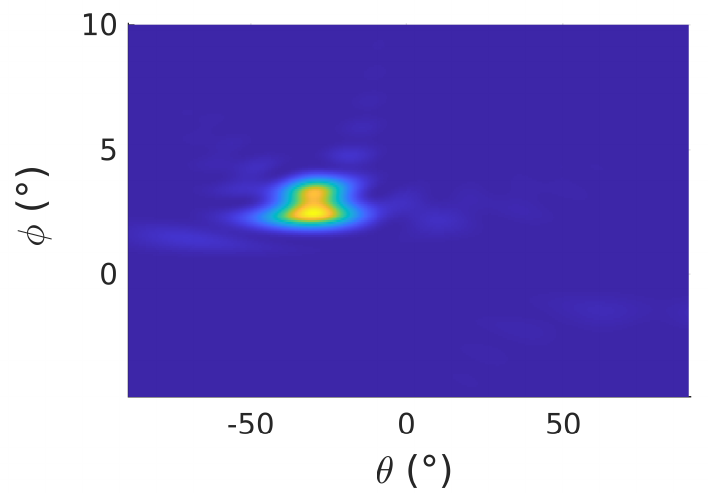}
    }
    \subfloat[OSCB-DOA: $ \Delta \phi = 1.1^{\circ}$]{
        \includegraphics[width=0.26\textwidth]{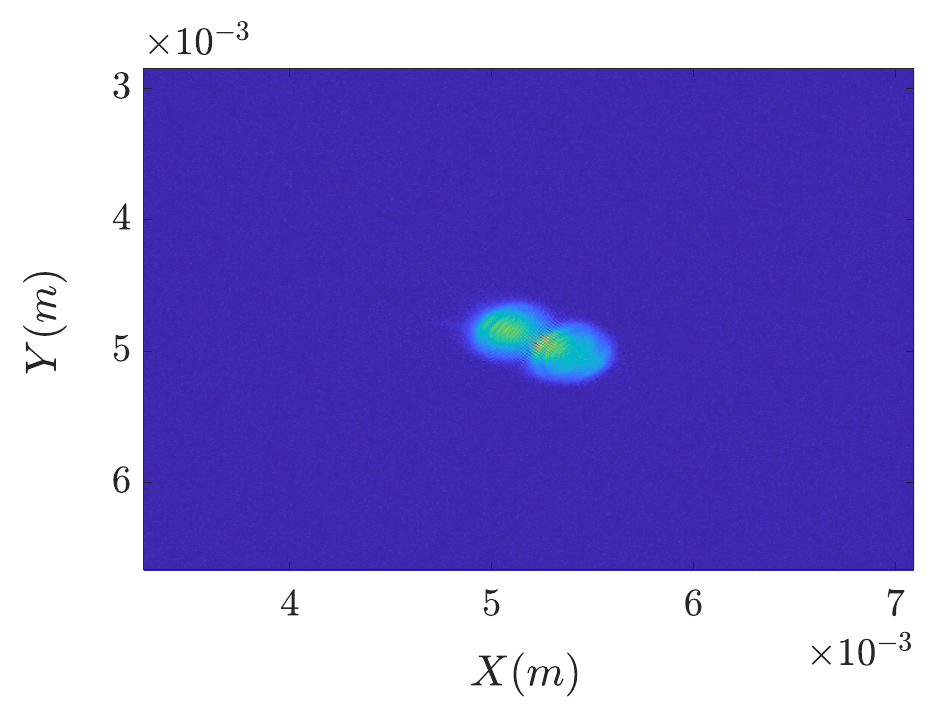}
    }
    \caption{Comparison of minimum resolving DOA angles for closely spaced sources using different algorithms.}
\end{figure*}
\begin{table}[htp]
    \centering
    \captionsetup{belowskip=0pt} 
    \caption{DOA estimation error under angular separations ($\Delta\phi$).}
    \begin{tabular}{lccccccc}
        \toprule
        DOA Separations ($^{\circ}$) & 1.10 & 0.90 & 0.70 & 0.50 & 0.30 & 0.1 \\
        \midrule
        OSB-DOA Error ($^{\circ}$) & 0.01 & 0.01 & 0.01 & 0.01 & 0.01 & 0.01  \\
        OWPB-DOA Error ($^{\circ}$) & 0.20 & 0.20 & 0.20 & 0.5  & inf  & inf  \\
        OSCB-DOA Error ($^{\circ}$) & 0.02 & inf  & inf  & inf  & inf  & inf  \\
        \bottomrule
    \end{tabular}
\end{table}

\normalsize {\color{blue}Fig. 12 illustrates the spectral response characteristics of a single Tx with multiple resonant beam links under different elevation angles $\phi$. As $\phi$ increases, the amplitude of the spectral peak gradually decreases.} This attenuation is attributed to the reduced transmission efficiency of the optical resonator, which leads to a decline in the output power at the receiver. {\color{blue}This trend is further validated in Fig.~13, which shows that both the transmission efficiency $V_r$ and the output power $P_{\text{out}}$ of a single Tx decrease as the elevation angle $\phi$ increases.} Notably, when $\phi$ increases from $6^{\circ}$ to $7^{\circ}$, the output power exhibits a sharp drop, which corresponds to the significant attenuation of the spectral peak observed in Fig. 12. When the value of $\phi$ exceeds $7^{\circ}$, $P_{\text{out}}$ reaches $0$~W, indicating that the resonant beam can no longer be established, which corresponds to the FoV being limited to $\pm 7^{\circ}$.

Since the OSB-DOA algorithm estimates the DOA by identifying spectral peaks, such power attenuation may cause the main peak of a weaker source to fall below the sidelobe level of the strongest frequency component of a stronger source, resulting in peak detection failure. Experimental results show that when the peak power ratio between different resonant beam sources exceeds 30, the weaker source cannot be reliably distinguished.

\begin{figure}[!t]
\centering
\includegraphics[width=2.8in]{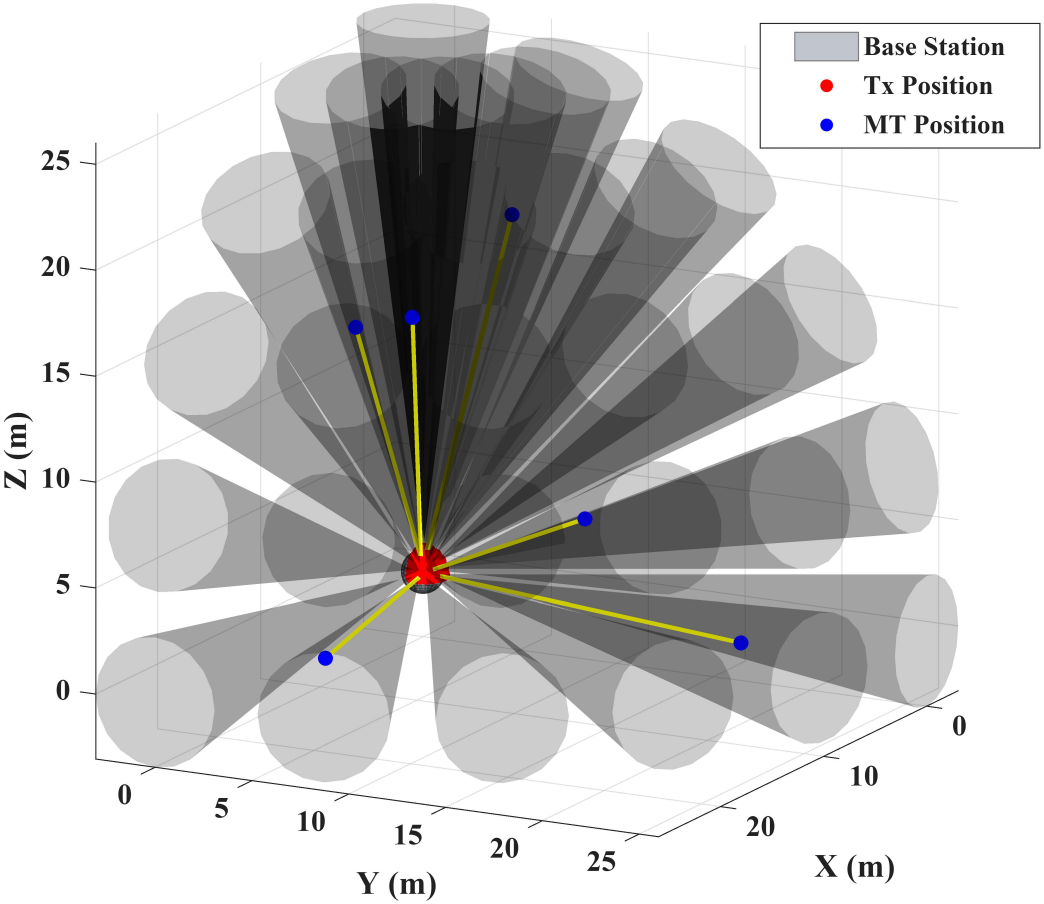}
\caption{{\color{blue}Multi-Tx integration and FoV expansion with overlapping coverage regions.}}
\end{figure}

Additionally, we evaluated the minimum resolvable angle performance of three {\color{blue}DOA} estimation algorithms suitable for resonant beams in multi-target scenarios. By gradually reducing the {\color{blue}elevation} angle spacing \(\Delta\phi\) between two sources, the corresponding DOA estimation errors were recorded (as shown in Table~II). When the two sources could not be distinguished, the result was marked as inf indicating that the estimation error tended toward infinity. Table~II shows that the OSB-DOA algorithm can resolve signals with $\Delta \phi = 0.1^{\circ}$, demonstrating superior angular resolution. In contrast, the resolution capability for OWPB-DOA and OSCB-DOA decreases as the minimum distinguishable angle increases to $0.5^{\circ}$ and $1^{\circ}$, respectively. These results indicate that the OSB-DOA algorithm offers a clear resolution advantage in multi-target scenarios, enabling reliable discrimination of closely spaced sources and accurate DOA estimation.

Fig.~14 further illustrates the energy spectrum distribution of each algorithm under the critical resolution condition. {\color{blue}Under this condition, the spectral peaks of multiple sources become closely spaced and eventually indistinguishable, leading to a significant degradation in DOA estimation accuracy.} As \(\Delta\phi\) decreases further, this effect becomes more pronounced, ultimately resulting in the failure of the algorithms to resolve the sources. This validates the analytical results from Section~{\color{blue}III-C}.

\subsection{Multi-Tx Integration Performance Analysis}

\begin{figure}[!t]
\centering
\includegraphics[width=2.8in]{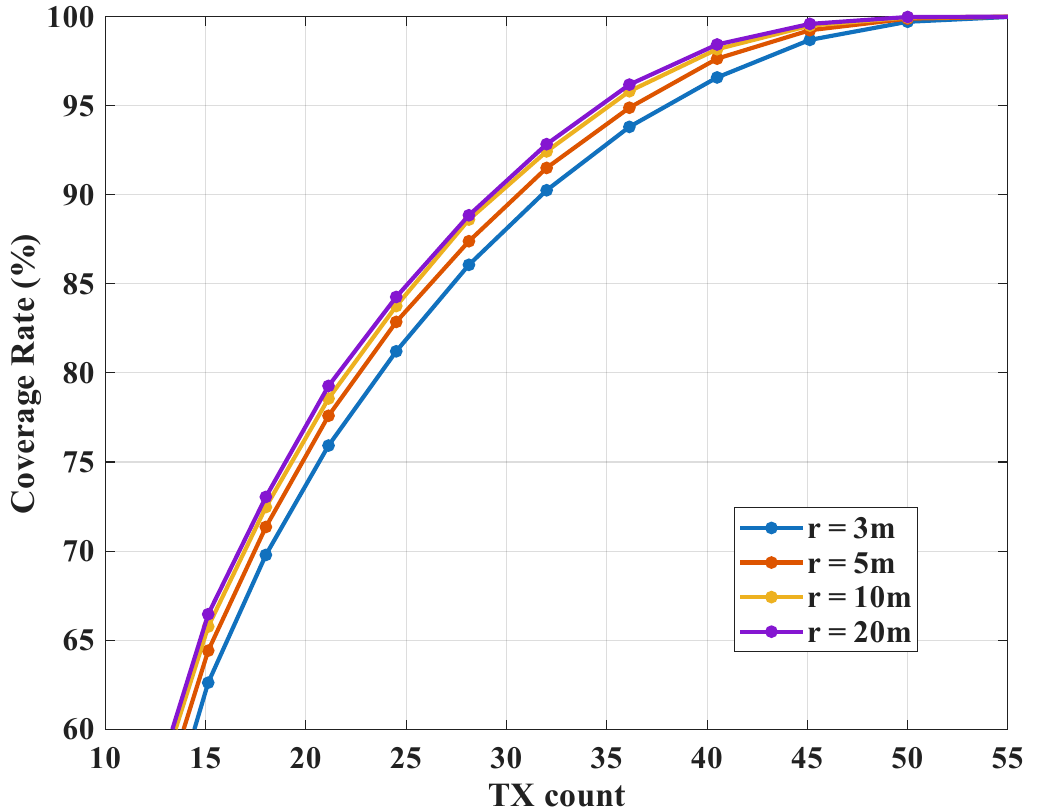}
\caption{Coverage rate versus Tx for different sphere radius (r).}
\end{figure}

Based on the foregoing analysis, the usable FoV of a single Tx device is $\pm 7^{\circ}$, which is limited in practical deployment. Therefore, the RB-HWDOA system incorporates a TM module to naturally integrate multiple Tx devices, thereby effectively expanding the FoV, as illustrated in Fig.~15. When multiple Tx devices are deployed on a spherical surface centered at the sensor, the FoV of each Tx overlaps, significantly increasing the system's coverage area. As the density of Tx devices increases, the FoV of the RB-HWDOA system gradually expands. Particularly in regions where multiple Tx FoVs overlap, the {\color{blue}uncovered} zones are markedly reduced, thus improving the operational range of the system.

To further validate the relationship between the number of Tx and the coverage rate of the RB-HWDOA system, as illustrated in Fig.~16, the Tx devices are uniformly distributed over an octant of a sphere of radius ${\color{blue}f_2 =} 0.15$ m. The coverage rate is evaluated over a concentric spherical octant of radius $ r $. The results indicate that the variation trends of the coverage rate across different detection radii are consistent as the number of Tx devices increases. When fewer than 15 Tx devices are deployed, the coverage rate increases rapidly, reaching approximately $60\%$. Beyond this point, further increases in the number of Tx devices lead to a gradually diminishing growth rate in coverage. The coverage rate eventually approaches $100\%$ when the number of Tx devices exceeds 50.

{\color{blue}Moreover, a larger test radius requires fewer Tx units to achieve the same coverage ratio. This is because, as illustrated in Fig.~15, each Tx covers a conically expanding region; therefore, evaluation on a larger spherical surface yields a higher per-Tx coverage ratio, reducing the total number of required Tx units.}

These findings demonstrate that increasing the number of Tx devices significantly improves both the FoV and the coverage range. However, the marginal improvement decreases beyond a certain number of devices. Thus, in practical deployments, the number and distribution of Tx devices should be optimized to achieve a balance between coverage and system efficiency. Moreover, for any given number of Tx devices, the coverage rate is consistently higher for larger detection radii than for smaller ones. This observation indirectly confirms the presence of overlapping FoV regions resulting from the integration of multiple Tx devices.

\section{Conclusion}
This paper proposes a high-resolution wide-{\color{blue}FoV} DOA estimation system named RB-HWDOA, based on resonant beam technology. The RB-HWDOA system introduces an optical spectrum-based DOA (OSB-DOA) estimation algorithm, which leverages the phase-sensitive oscillation of resonant beams to overcome the beam width-limited resolution of amplitude-only spatial distributions and achieves high-resolution DOA estimation in multi-target scenarios. By integrating the TM structure into the Tx, the proposed RB-HWDOA achieves a multi-Tx integrated architecture, effectively extending the {\color{blue}FoV}.

Simulation analysis demonstrates that the OSB-DOA algorithm significantly enhances angular resolution while maintaining robustness to noise and low computational complexity. Unlike conventional DOA estimation methods, the OSB-DOA approach achieves a minimum distinguishable angle of $0.1^{\circ}$, exhibiting excellent performance in distinguishing closely spaced {\color{blue}multiple targets}. Furthermore, the integration of multiple Tx devices via the TM module substantially extends the {\color{blue}FoV}, facilitating the deployment of the RB-HWDOA system in large-scale or complex environments. By combining the advantages of resonant beam technology, RB-HWDOA provides a scalable DOA estimation solution for IoT applications, {\color{blue}with potential for deployment in scenarios such as autonomous driving and smart homes}.

{\color{blue}Future work will focus on establishing reliable non-line-of-sight links and developing robust DOA sensing methods for non-stationary resonant beams to facilitate practical deployment of RB-HWDOA in IoT environments.}

\end{document}